\documentclass[aps,preprint,superscriptaddress]{revtex4}

\usepackage[english]{babel}
\usepackage{braket}
\usepackage{epsfig}
\usepackage{graphics}
\usepackage{graphicx}
\usepackage{xcolor}
\usepackage{hyperref}
\usepackage{natbib}
\usepackage{amsmath}
\usepackage{times}
\usepackage[T1]{fontenc}
\usepackage[utf8]{inputenc}
\usepackage{ulem}
\usepackage{psfrag}
\usepackage{subfigure}
\usepackage{epstopdf}

\usepackage[toc]{appendix}
\begin{document}

\title{

Stability analysis of the  Hindmarsh-Rose neuron under electromagnetic induction}
 
%
\author{\textbf{L. Messee Goulefack}}
\affiliation{Fundamental Physics Laboratory, Physics of Complex Systems group,
Department of Physics, Faculty of
 Science, University of Douala, Box 24 157 Douala, Cameroon}
 \author{\textbf{A. Cheage Chamgoue}}
 \affiliation{Department of Basic Science, School of Geology and Mining Engineering,
University of Ngaoundere, Box 115, Meiganga, Cameroon}
\author{\textbf{C. Anteneodo}}
\affiliation{Departamento de F\'isica, Pontif\'icia Universidade Cat\'olica do Rio de Janeiro, \& National Institute of Science and Technology (INCT) of Complex Systems, Rua Marqu\^es de S$\tilde{a}$o Vicente, 225--22451--900 G\'avea - Rio de Janeiro -Brazil }

\author{\textbf{R. Yamapi}}
\email[ryamapi@yahoo.fr]{(Corresponding authors)}
\affiliation{Fundamental Physics Laboratory, Physics of Complex System group,
Department of Physics, Faculty of
 Science, University of Douala, Box 24 157 Douala, Cameroon}
\date{\today}

\begin{abstract}
We consider the Hindmarsh-Rose neuron model modified by taking into account the effect of electromagnetic induction on membrane potential. We study the impact of the magnetic flux on the neuron dynamics, through the analysis of the stability of fixed points. 
Increasing magnetic flux  reduces the number of equilibrium points and favors their stability. 
Therefore, electromagnetic induction tends to regularize chaotic regimes and  to affect  regular and quasi-regular ones by reducing the number of spikes or even destroying the oscillations.  
 

\end{abstract}
\maketitle


\section{Introduction}

The functioning of the fundamental cell of the nervous system
(the neuron) has been the subject of research in  various scientific fields. Beyond the fundamental questions of neuroscience, advances in understanding the mechanisms of neuronal activities and their responses to external stimuli can be  helpful, for instance, in the development of artificial intelligence and other technologies.
So,  many efforts have been made  
  to grasp, through  
   model systems, the dynamics of real biological neurons. 
   Among the successful mathematical models, consistent with experimental observations, let us mention those of  Hodgkin-Huxley~\cite{hodgkin52,hasegawa00,kang15},
   Morris-Lecar~\cite{morris81,hu16,shi14}, Izhikevich~\cite{izhikevich03}, FitzHugh-Nagumo~\cite{kosmidis03,abbasian13,wang07}
   and Hindmarsh-Rose (HR)~\cite{HR82,HR84,innocenti07,wu16}, to name just a few.
These models can be improved by incorporating specific features.  
For instance in neural circuits, the introduction of piezoeletric  ~\cite{auditory2021} or light-sensitive elements~\cite{light2021}  is currently under investigation  to mimic auditory or visual responses. 
 
Another important feature that gained attention more recently is the influence of induced magnetic fields. 
Electrophysiological activity 
can induce magnetic fluxes due to time-varying currents, which can affect the membrane potential. Moreover, although electrical and chemical synapses have a crucial role in the transmission of information between neurons,  the exchange of signals can also occur through the flow of ionic currents through gap junctions that allow direct passage between cells, a mechanism that can be affected by  the influence of electromagnetic fields. 
The effect of electromagnetic induction has been taken into account introducing modifications in the above mentioned neuronal models. %
This has typically been done via memory resistance (memristor) coupling of the magnetic flux to membrane potential~\cite{bao10,muthuswamy10,li15},  
in Fitzhugh-Nagumo~\cite{wuwang16},   Hodgkin-Huxley~\cite{wuwang17,xu18}  and HR~\cite{lv16,wuxu17,parastesh18,usha19} neurons. 
The inclusion of memristive effects has been shown crucial, for instance, to explain effects in heart tissues due to electromagnetic radiation~\cite{wuwang16}.
Networks of coupled neurons, under magnetic flow, have also been investigated~\cite{parastesh18,parastesh19,mostaghimi19,alternating}, identifying diverse spatiotemporal patterns including wave propagation and chimera states. The later are particularly interesting since are intermediate between order and disorder and have been observed in diverse neural networks~\cite{lakshmanan2019,makarov2021}. 

In this work we focus on the single neuron dynamics with the inclusion of memristive effects. 
We consider an extended version of the HR neuronal model, previously proposed to take into account the adjustment of the membrane potential due to a magnetic flux across it~\cite{lv16}. This model has been investigated before,  mainly for oscillatory   external current~\cite{lv16} or  external field~\cite{parastesh18}, or   both~\cite{wuxu17}. 
Here we study  the dynamical regimes that arise when changing the strength of the induction coupling, under constant external inputs. 
The type of regime is relevant since it can determine the neuro-computational capacity of the cells to process the inputs and communicate the output to other cells~\cite{morris81,houart99,izhikevich00}.

The paper is structured as follows. In Section~\ref{sec:model}, we summarize the MHR model  and define the values of the parameters.
In Section ~\ref{sec:dynamics}, we  present the results from the stability analysis and  bifurcation diagrams. 
Section \ref{sec:conclusions}, we highlight the main findings.

\section{Neuronal model of Hindmarsh-Rose with electromagnetic induction}
\label{sec:model}

The  HR model of  neuronal bursting has been proposed in Ref.~\cite{HR84} and thereafter has been intensively studied and extended in several directions~\cite{houart99,innocenti07,yamapi13,wang13,hilda16,wu16,rajagopal19}. 
In its original version, it consists of  the following  set of three coupled first-order nonlinear differential equations: 
\begin{equation}
\label{eq:HR0}
\left\{
\begin{array}{ll}
\dot{x}=y-ax^{3}+bx^{2}-z+ I_{ext},\\
\dot{y}=c-dx^{2}-y,\\
\dot{z}=r[s(x-x_{0})-z],
\end{array}
\right.
\end{equation}
where the variables $x$, $y$ and $z$ describe the membrane potential,
the recovery and  adaptation  ionic currents, respectively,
with $I_{ext}$  denoting the external forcing current, and $a$, $b$, $c$, $d$, $r$, $s$ and $x_0$
are typically positive constants. 
As expected for biological neuron models, it takes into account the membrane potential as well as the currents through ion channels that regulate  ion propagation. 
However,  neuronal activity depends on complex external and internal influences. In particular, the effect of the  magnetic flux $\phi$ across the cell membrane can affect its potential via a memristor effect.  
Then, to describe the interaction between neuronal activity and a magnetic flux,
  a fourth dimension was added to Eq.~(\ref{eq:HR0}), and the resulting four-dimensional neuron model,  modified HR (MHR)~\cite{lv16,usha19}, is expressed as 
\begin{equation}
\label{eq:MHR}
\left\{
\begin{array}{ll}
\dot{x}=y-ax^{3}+bx^{2}-z+ I_{ext}-kxM(\phi),\\
\dot{y}=c-dx^{2}-y,\\
\dot{z}=r[s(x-x_0)-z], \\
\dot{\phi}=k_{1}x-k_{2}\phi,
\end{array}
\right.
\end{equation}
where
$ M(\phi)$ represents the coupling between the magnetic flux across the membrane, $\Phi$, and the membrane potential $x$. 
The term $-k M(\Phi)x$ denotes the current produced through electromagnetic induction, modulated by the intensity $k$. This current, together with the external current  $I_{ext}$, contributes to the change in the membrane potential (see Ref. \cite{lv16} for further details). 
$M(\phi)$ is assumed equivalent to the memductance of a magnetic flux-controlled memristor~\cite{li15,lv16},
modelled by 
  $ M(\phi)= \alpha + 3\beta \phi^{2} $~\cite{lv16,usha19},  where  $ \alpha$ and  $ \beta $ are positive parameters. 
 This quadratic form  represents a minimal nonlinear model, smooth,  positive definite, increasing with the flux intensity. Let us mention that discontinuous (piece-wise linear) forms have been also studied in the literature of the modified HR neuron~\cite{parastesh18}.  
Finally, the parameters   $k_{1}$ and $k_{2}$ are rates that control the evolution of the magnetic flux, governed by the membrane potential and  leakage.
 
 Let us remark that 
other extensions of the HR neuron have been considered before, for instance with variable signals~\cite{lv16,wuxu17} in the MHR, or 
 modification of the equation of the recovery current in the HR~\cite{hilda16,rajagopal19}, differently to the MHR model here considered, in which the equation for the potential is adjusted.
 
In the numerical simulations, we will use the physiologically relevant values
$a=1$, $b=3$, $c=1$, $d=5$, as in the standard~\cite{HR84} and extended~\cite{lv16} HR models. 
The threshold potential was set $x_{0}= -1.6$, and we will 
typically set $r=0.001$ and $s=4$, unless other values are specified.  
For the memory function, we set $\alpha=0.1$ and  $\beta =0.06$, and the coefficients that rule the magnetic flux dynamics are
$k_{1}=0.1$, $k_{2}=0.5$~\cite{yamapi13}.

The constant external excitation current $I_{ext}$ and the strength $k$ of the magnetic term are the main control parameters, which will be varied throughout this work. In stability  analyses, we will also vary the adaptation parameters $r$ and $s$.

\section{Analysis of the dynamics of the MHR model}
\label{sec:dynamics}

\subsection{Equilibrium points}
\label{sec:equilibria}

\begin{figure}[b!]
\includegraphics[scale=0.7]{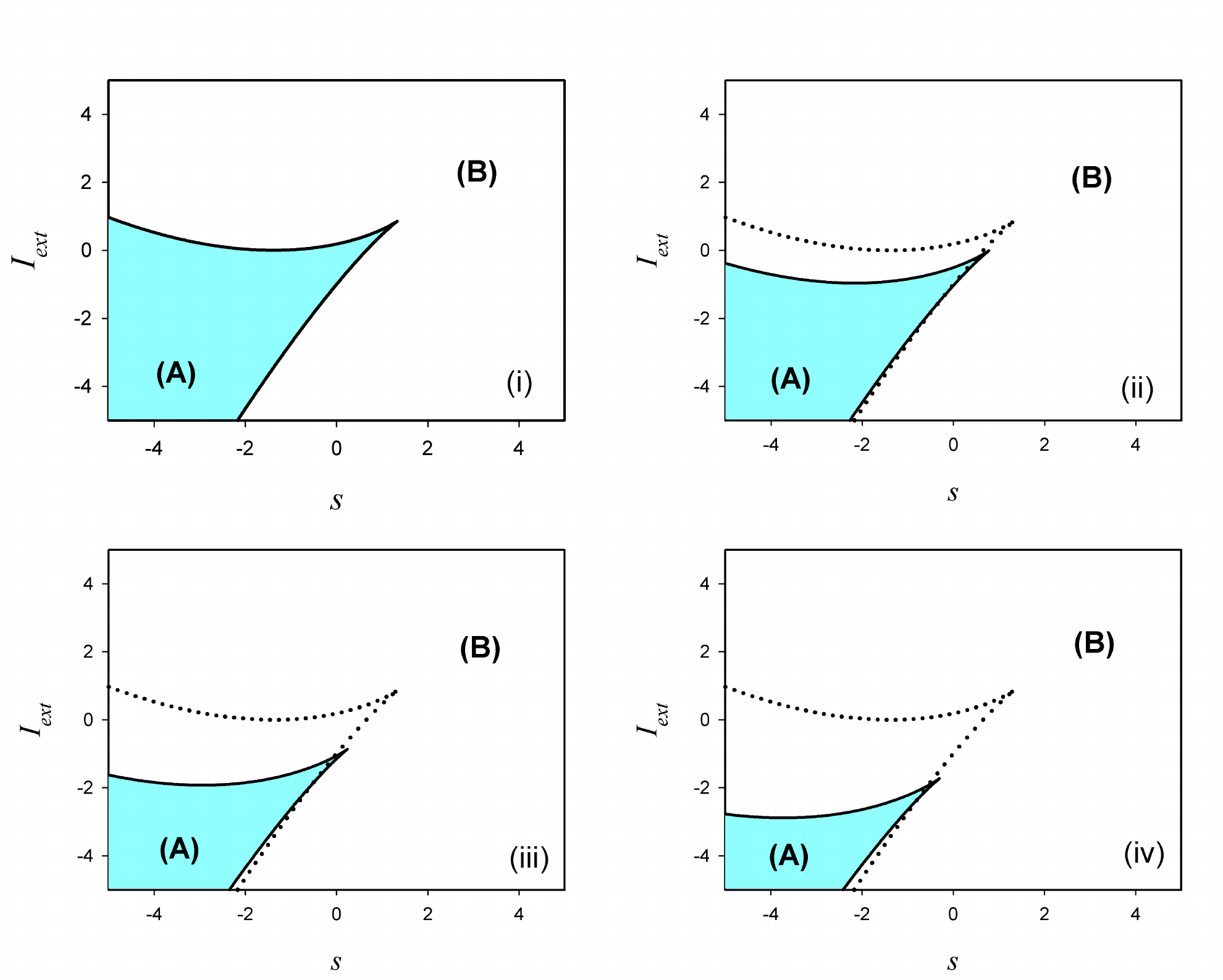}
\caption{ Regions characterized by the number of equilibrium points in the $s-I_{ext}$ plane, for different values of the magnetic-flux coupling parameter: 
 $k=0$ (i), $k=5$ (ii), $k=10$ (iii) and  $k=15$ (iv).
In the shadowed region (A) there are three equilibrium points, while in region (B), 
 there is a single equilibrium point. At the border (full lines), there are two equilibrium points. The dotted line repeats the case $k=0$, for comparison.
Notice that the region (A) shrinks with increasing $k$.    
}
\label{fig:sI}
\end{figure}

The fixed points of the system of equations presented in Eq.~(\ref{eq:MHR})  are obtained by setting the time derivatives equal to zero, which leads to  a systems of nonlinear algebraic equations whose solution gives  equilibrium points of the form~\cite{parastesh18}
   \begin{equation} 
       E=(x_{e},y_{e},z_{e},\phi_{e})=(x_{e}, -dx_{e}^{2}+c, s(x_{e}-x_{0}), k_{1}x_{e}/k_{2}),  
   \end{equation}
where the equilibrium potential $ x_{e}$  satisfies the equation 
\begin{equation}
\label{eq:xe}
a_{0}x_{e}^{3} + a_{1}x_{e}^{2} + a_{2}x_{e} + a_{3} = 0,
\end{equation}
with the expressions for the coefficient $ a_{i}$ $(i= 0, 1, 2, 3) $ given by
\begin{equation}
\label{eq:as}
\left\{
\begin{array}{ll}
a_{0}=-\left(a+ \dfrac{  3 k\beta k_{1}^{2}}{k_{2}^{2}} \right)   \equiv -T,  \\
a_{1}= (b-d),\\
a_{2}= -(s+ k\alpha),\\
a_{3}= sx_{0} + I_{ext} + c.
\end{array}
\right.
\end{equation}

The number of real roots of Eq.~(\ref{eq:xe}) depends on  the sign of its discriminant $\Delta$ 
 defined in Appendix \ref{appa}  (see Eq.~(\ref{eq:discriminant})).
As a consequence, the neuronal model has 
three equilibrium points for  $\Delta<0$ (defined by $x_e$ given in  Eq.~(\ref{eq:E3})), 
one equilibrium point $E$
for $\Delta>0$  (with $x_e$ defined in Eq.~(\ref{eq:E1})),
while at the boundary where 
 $\Delta=0$, there are 
 two equilibrium points (with $x_e$ defined in Eq.~(\ref{eq:E2})). 
This boundary is explicitly given by  
\begin{eqnarray}	I^\pm_{ext}
&=&- (c+sx_0) + \frac{ 9T(b-d)(s+k\alpha)-2(b-d)^3}{27\,T^2} \pm  \frac{2\bigl[   (b-d)^2-3T (s+k\alpha)\bigr]^{3/2}}{27\,T^2},\;\;\;
\label{eq:Ipm}
\end{eqnarray}
which defines two lines in the $s-I_{ext}$ plane, plotted in Fig.~\ref{fig:sI}, for different values of $k$. Between the two  curves, region (A) corresponding to $\Delta<0$,
there are three equilibrium points, 
in the boundary curves, there are two 
equilibrium points, while in the region (B) corresponding to $\Delta>0$, the neural model has 
a single equilibrium point.
The case without magnetic flux has been reported before~\cite{yamapi13}. Now we extend that result to 
show the effects of the electromagnetic induction on the boundaries of the stability regions by varying the magnetic coupling strength $k$. 
Notice that, as  $k$ increases, from  $k = 0$ (in Fig.~\ref{fig:sI}.i)
to $k = 15$ (in Fig.~\ref{fig:sI}.iv), 
the domain of existence of three equilibrium points is reduced. 
The analysis of the conditions under which these points are stable or unstable is developed in Sec.~\ref{sec:stability}.

\subsection{Stability analysis}
\label{sec:stability}

To study the linear stability of equilibrium points, let us introduce the deviation vector

\begin{eqnarray}
\label{eq:deltas}
\delta  X=(\delta   x,\delta   y,\delta  z,\delta \phi)^T =
(x-x_{e}, y-y_{e}, z-z_{e},  \phi-\phi_{e})^T,
\end{eqnarray}
which  measures  the nearness between 
a dynamical state $X=(x,y,z,\phi)$
and the equilibrium point $E=(x_e,y_e,z_e,\phi_e)$. 
Linearization of  Eq.~(\ref{eq:MHR}) leads to
\begin{equation}
\label{eq7}
\delta \dot X = J(x_{e},y_{e},z_{e},\phi_{e}) \, \delta X,
\end{equation}
where $\delta \dot X=(\delta \dot x,\delta \dot y,\delta \dot z,\delta \dot \phi)^T $,
%
and $ J (x_{e},y_{e},z_{e},\phi_{e})$ is the Jacobian matrix  of system (\ref{eq:MHR}) around the equilibrium point
$(x_e,y_e,z_e,\phi_e)$,
namely,
\begin{equation}
\label{eq:J}
		 J(x_{e},y_{e},z_{e},\phi_{e}) =\begin{bmatrix}
		-3ax_{e}^{2}+2bx_{e}-k(\alpha + \dfrac{3\beta  k_{1}^{2}}{k_{2}^{2}}x_{e}^{2})& 1 & -1 & - \dfrac{6k\beta  k_{1}}{k_{2}}x_{e}^{2}\\
		-2dx_{e}& -1 & 0 & 0\\
		rs& 0 & -r & 0\\
		k_{1}& 0 & 0 & -k_{2}
		\end{bmatrix} .
		\end{equation}
The linear stability of the equilibrium  states
is given by the eigenvalues $\lambda$ of the Jacobian matrix $J$.
If the real parts of the
roots of the resulting characteristic equation  are all negative, the corresponding
equilibrium states are stable. If at least one root has a positive real part,
the equilibrium states are unstable. 

The characteristic equation of the Jacobian matrix is   
\begin{equation}
\label{eq:deltai}
\lambda ^{4} +\delta_{1}\lambda ^{3} + \delta_{2}\lambda ^{2} +\delta_{3}\lambda +\delta _{4}=0, 
\end{equation}
where, the coefficients $\delta_i$ are explicitly given in the Appendix \ref{appb}. 
The determination of the sign  of the real part  of
the roots $\lambda$ may be carried out by making use of the Routh–Hurwitz stability criterion~\cite{hayashi64}.
According to  this criterion,  in our case, the real parts of  all the roots of the characteristic polynomial are negative whenever
 \begin{equation}
 \label{eq:characteristic}
 \left\{
  \begin{array}{ll}
  \delta_{i}>0, \;\;\;\mbox{for all $i=1, 2, 3, 4$},\\
  \delta_{1}\delta_{2}\delta_{3}> \delta_{3}^{2} +\delta_{1}^{2}\delta_{4}. 
  \end{array}
   \right.
 \end{equation}
 
\begin{figure}[h!]
 \includegraphics[width=6.3cm,height=4.7cm]{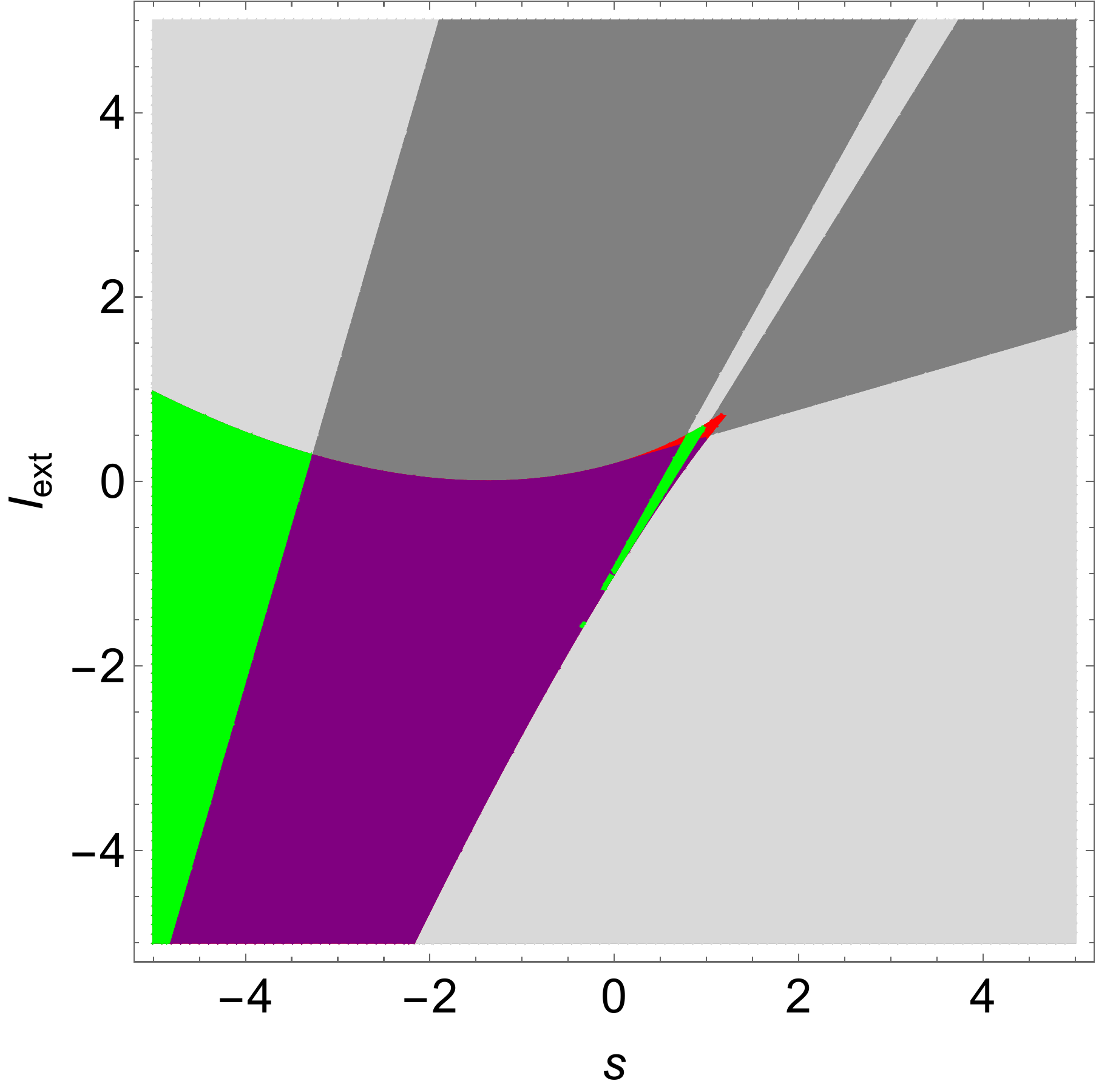}(i)$\;\,$
 \includegraphics[width=6.3cm,height=4.7cm]{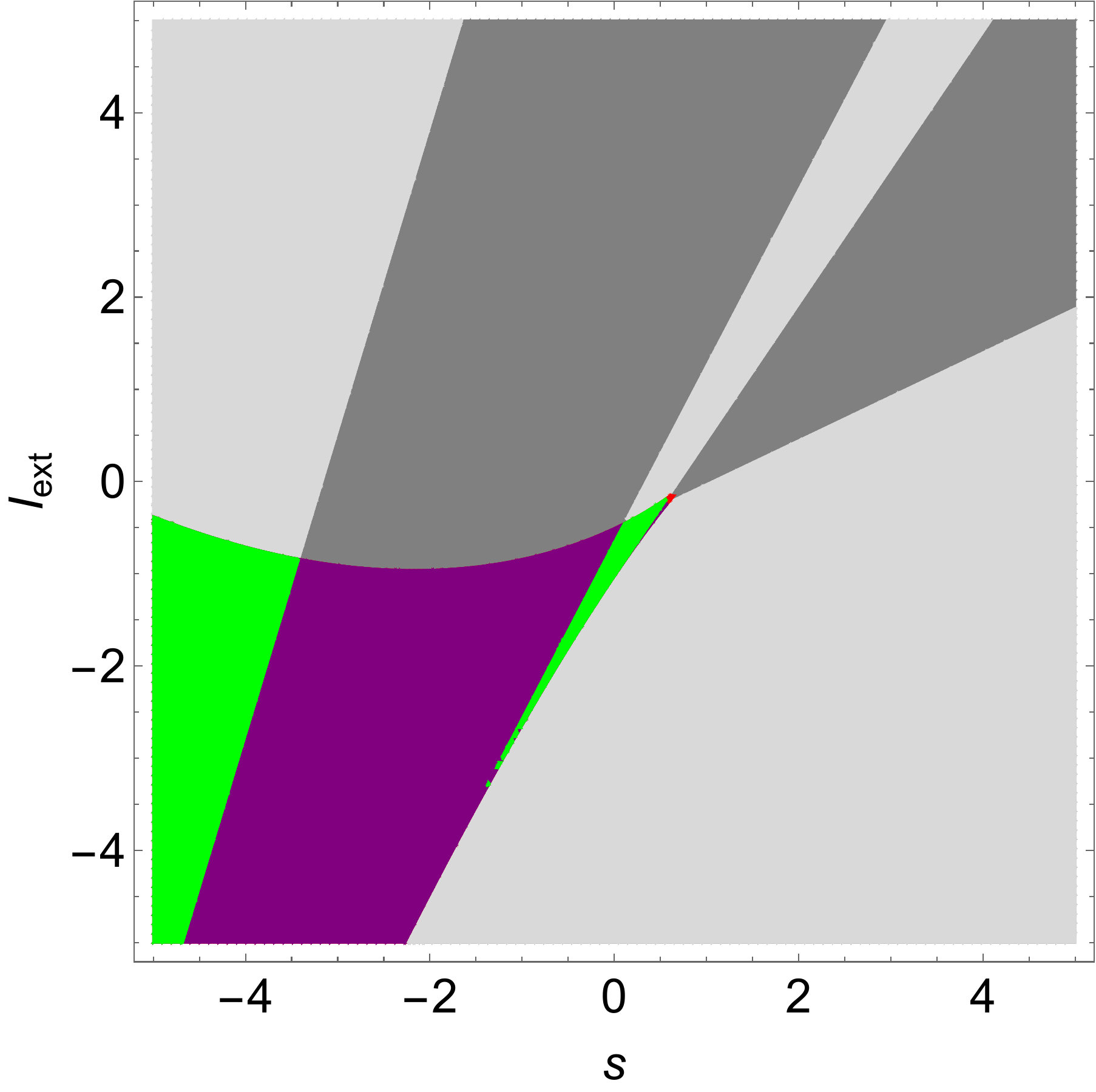}(ii)
\\
 \includegraphics[width=6.3cm,height=4.7cm]{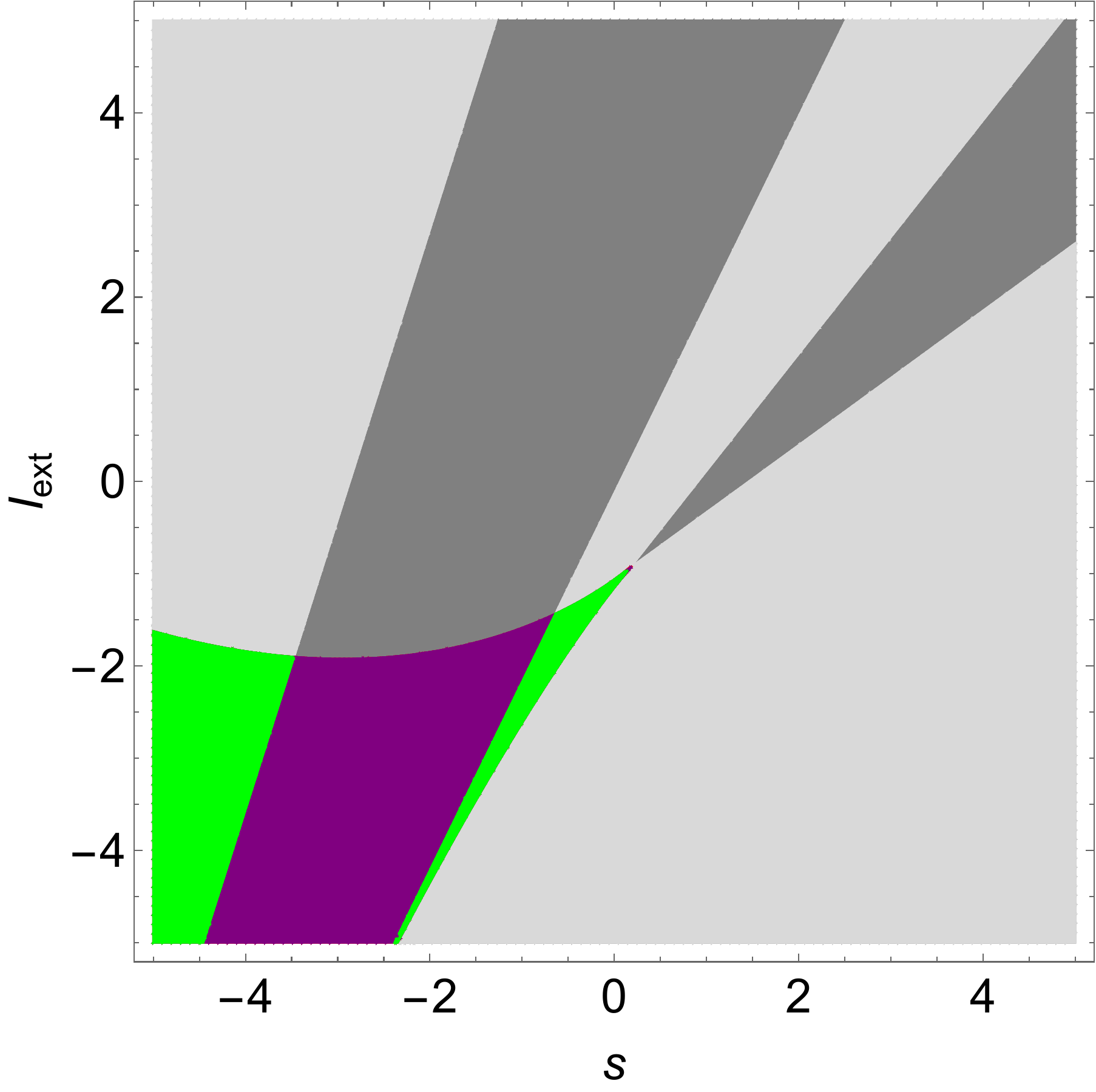}(iii)
 \includegraphics[width=6.3cm,height=4.7cm]{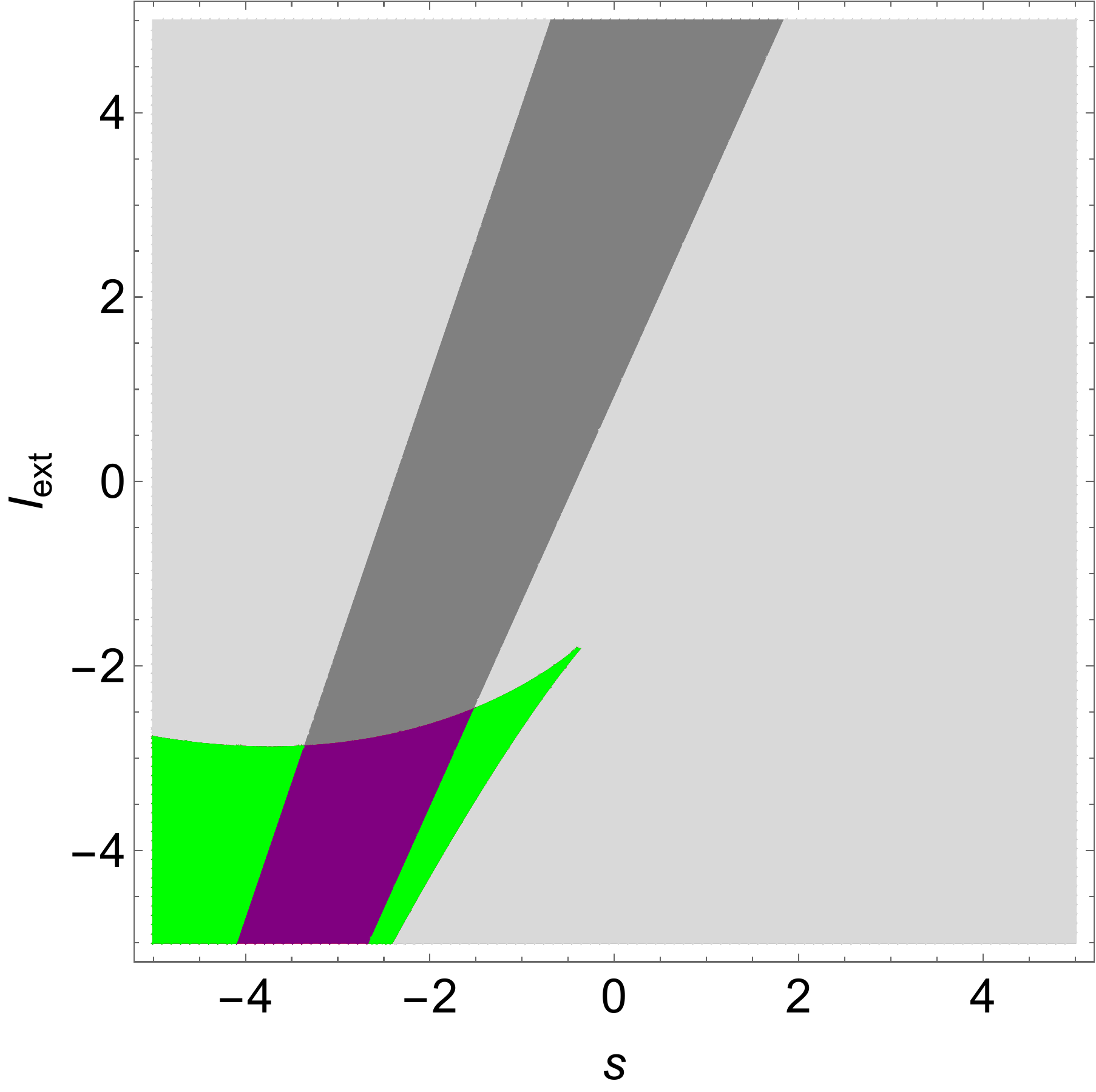}(iv)
 \caption{Stability maps of the equilibrium points $E$, in the plane $s-I_{ext}$,  for $k=0$ (i), $k=5$ (ii), $k=10$ (iii) and  $k=15$ (iv). 
The colored regions correspond to 
one equilibrium point [unstable (dark-gray),
  stable (light-gray)] and  
  three equilibrium points [1 unstable (green), 2 unstable (purple), all unstable (red)]. 
  The latter only appears, within the resolution of the figure, near the vertex of region (A).
 }
 \label{fig:nature}
 \end{figure}

We have already shown the regions defined by the number of equilibrium points in the plane  
$s-I_{ext}$  in Fig.~\ref{fig:sI}, noticing a considerable modification of the boundaries of these regions when increasing the magnetic coupling $k$.   
Now we present, in Fig.~\ref{fig:nature}, the subdomains in the $s-I_{ext}$ plane where the equilibrium points have different  stability, focusing on the effects of  electromagnetic induction.

Let us start by analyzing the case where
the neuronal model is not subjected to any  magnetic flux (i.e., $k = 0$), which is depicted in  Fig.~\ref{fig:nature}.i). 
In region (B), we distinguish the  subregions where the single equilibrium point is unstable (dark-gray) or stable (light-gray),  
Region (A) is mainly composed of two subdomains (green and purple), where, among the three equilibrium points defined by $(s,I_{ext})$,  
only one is unstable (green), or two are unstable (purple). A small subdomain (red) where the the three equilibrium points are all unstable is also observed near the vertex of region (A). 
 
In region (B), a stable fixed point means extinction of the oscillations. Taking into account the electromagnetic induction, with intensity $k$, clearly the region (B) gains stability with increasing  $k$, as can be observed by the expansion of the light-gray region, in the successive panels of Fig.~\ref{fig:nature}. 

Besides the reduction of  region (A), when $k$ increases, there is also a gain of stability, as can be seen by the  predominance  of the green subdomain over the purple one, and disappearance of the small red subdomain associated to three unstable fixed points. 
 In short, the progressive increase
 of  magnetic coupling tends to reduce the number of equilibrium points form 3 to 1 and turns the single equilibrium point more stable. Recall that $r=0.001$ is used, but a similar portrait to that shown in Fig.~\ref{fig:nature} is observed for other values of $0<r\lesssim 0.1$.

Illustrative examples are provided in the tables of Appendix \ref{appc}, where, for selected points $P=(s, I_{ext})$, 
we present the corresponding equilibrium states (whose number depends on the region which $P$ belongs to) and their corresponding 
eigenvalues, which express the nature of the fixed points. For comparison, for each point $P$, results are given for the 
neuronal MHR system with $k = 0$  and   $k=10$.
 

\subsection{Bifurcation diagrams}

We present, in this section,  bifurcation diagrams
in the MHR model, as a function of the control parameters. 
Trajectories were obtained  solving numerically Eq.~(\ref{eq:MHR}), 
 using a $4th$-order Runge-Kutta algorithm, typically starting from the initial condition 
$(x,y,z,\phi)_{t=0}=(0,0,0,0)$, and performing measurements in the interval 
$t \in (1000,8000)$. 
We also computed the largest Lyapunov exponent, defined by
\begin{equation}\label{eq:lmax}
  L_{max}=\lim_
  {  t\to \infty  \atop \delta {X}(0)\to 0 }
  \frac{1}{t}\ln\left|\frac{\delta {X}(t)}{\delta {X}(0)}\right|,
\end{equation}
 where ${X}$ is the state of the system and in this case
$\delta  {X}$ is the separation vector between two close trajectories. 
$L_{max}$ is estimated using Benettin algorithm, after  solving numerically Eq.~(\ref{eq:MHR}) and its associated variational equation. 

\begin{figure}[b!]
\includegraphics[width=16cm, height=8cm]{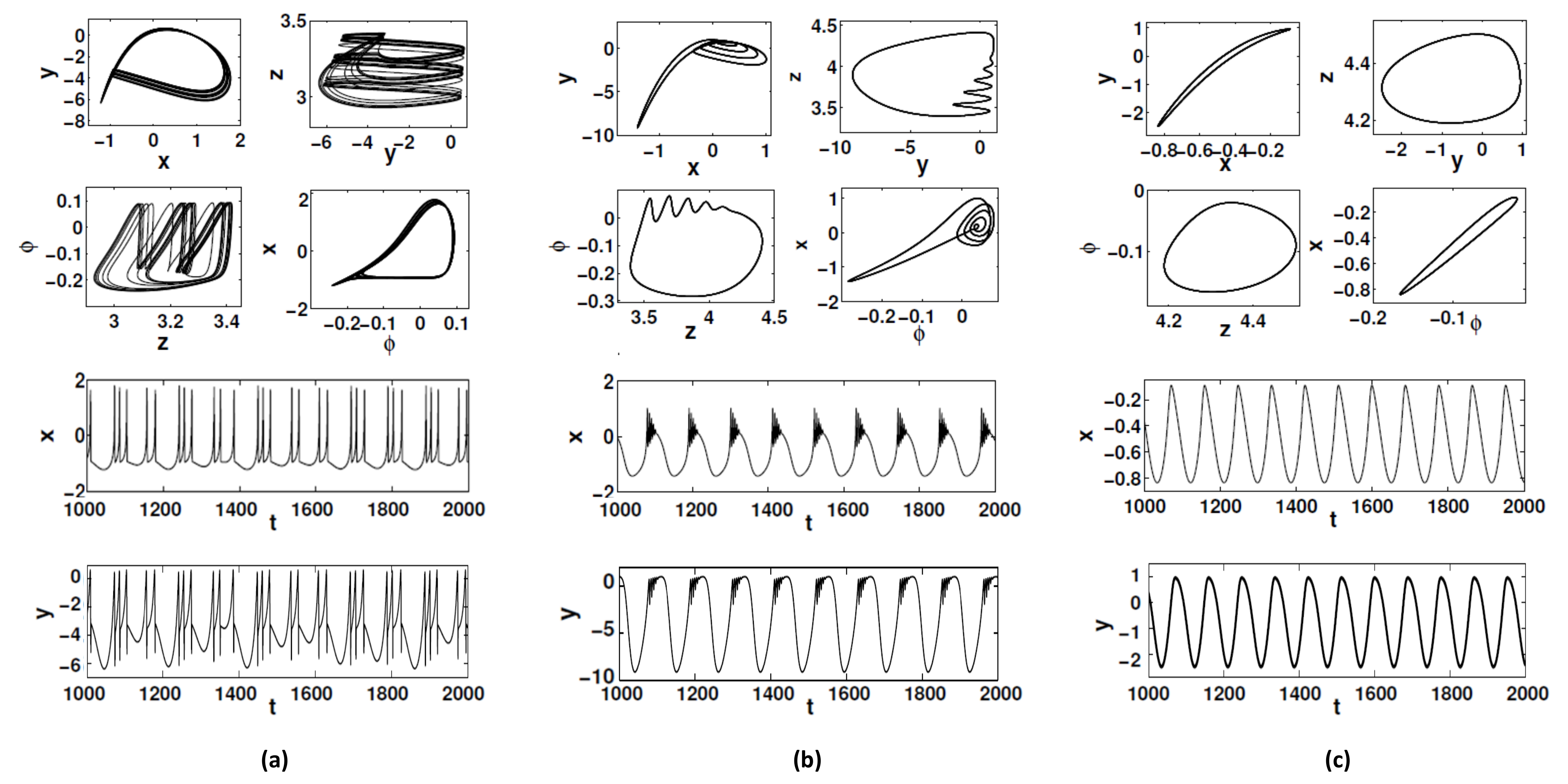} 
	\caption{ 
	Projections of the phase portrait in different planes, and corresponding time series for the membrane potential $x$ and the current $y$ vs. $t$, for (a) $k=0$, (b) $k=5$ and (c) $k=10$, with $r = 0.008$,  $s=4$ and $I_{ext}=3.25$. In (a), we observe the chaotic attractor of the HR neuronal model, while the dynamics becomes regularized as $k$ increases.
	}
\label{fig:attractor}
\end{figure}

Figure \ref{fig:attractor} shows 
the  projections of the phase portrait in different planes, as well as the time series for the membrane potential $x$, and the recovery current $y$, for different values of $k$.   
When $k=0$, the well-known chaotic attractor of the neuronal HR model~\cite{yamapi13,parastesh18,rajagopal19} is recovered (Fig.~\ref{fig:attractor}a), while as $k$ increases, the dynamics becomes more and more regular, the amplitude of the spikes per burst is reduced (e.g., Fig.~\ref{fig:attractor}b) and disappears for large enough $k$ (e.g., Fig.~\ref{fig:attractor}c).
The change in patterns can affect the performance of the neuron, impacting on information processing and transmission.

\begin{figure}[h!]
\centering
\includegraphics[width=8cm,height=6cm]{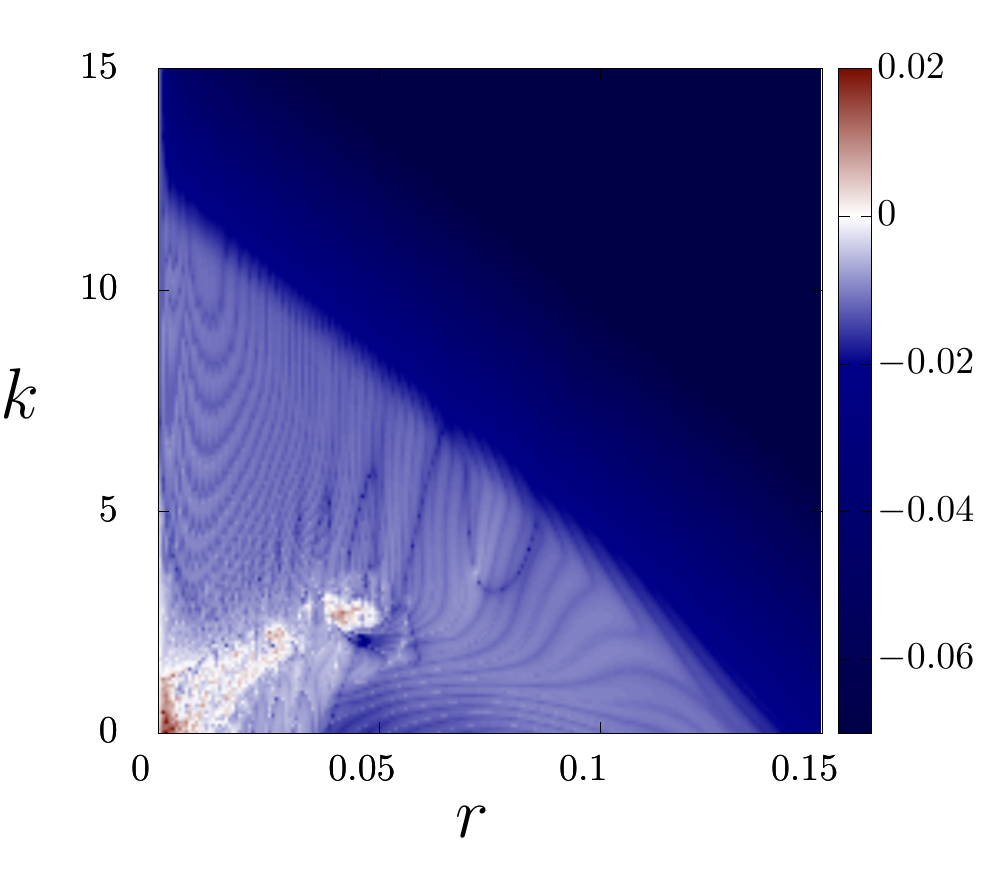} 
\includegraphics[width=8cm,height=6cm]{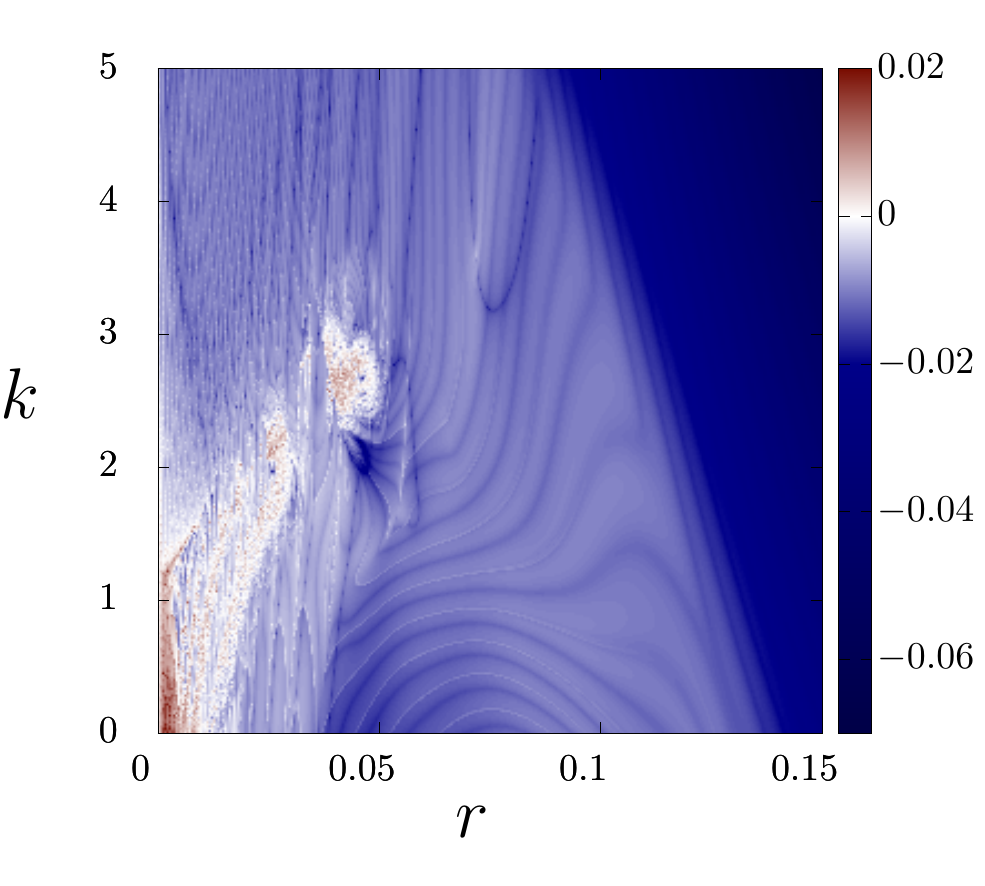} 
\caption{Heat-plots of the largest
Lyapunov exponent $L_{max}$ as a function of $r$ and $k$, using $s=4$, $I_{ext}= 3.25$. In the right-hand-side panel we amplified the low $k$ region. 
We observe that for sufficiently low $k$ chaotic trajectories (red scale) can exist when $r$ is also low, but increasing $k$ leads to negative values of $L_{max}$,  indicating regular behavior. The darker blue region corresponds to damped oscillations towards a fixed point.
}
\label{fig:kr}
\end{figure}

In Fig.~\ref{fig:kr} we present a heat-plot of the largest Lyapunov exponent in the plane $r-k$, which provides a full picture of the effects of   magnetic induction on the dynamics of the MHR,  illustrated for particular values of the strength $k$ in Fig.~\ref{fig:attractor}.

\begin{figure}[h!]
\centering
\includegraphics[width=8cm, height=5cm]{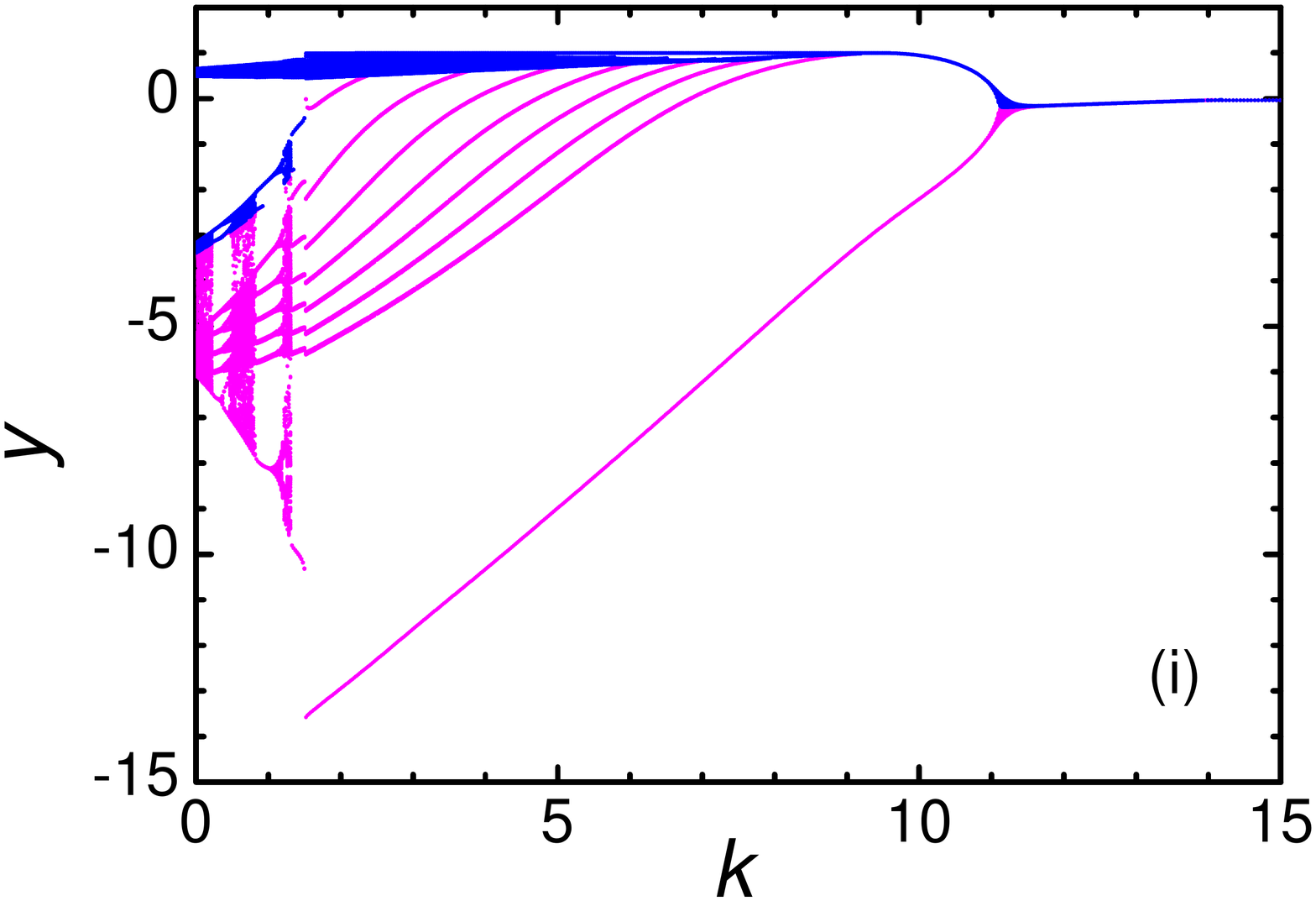}
\includegraphics[width=8cm, height=5cm]{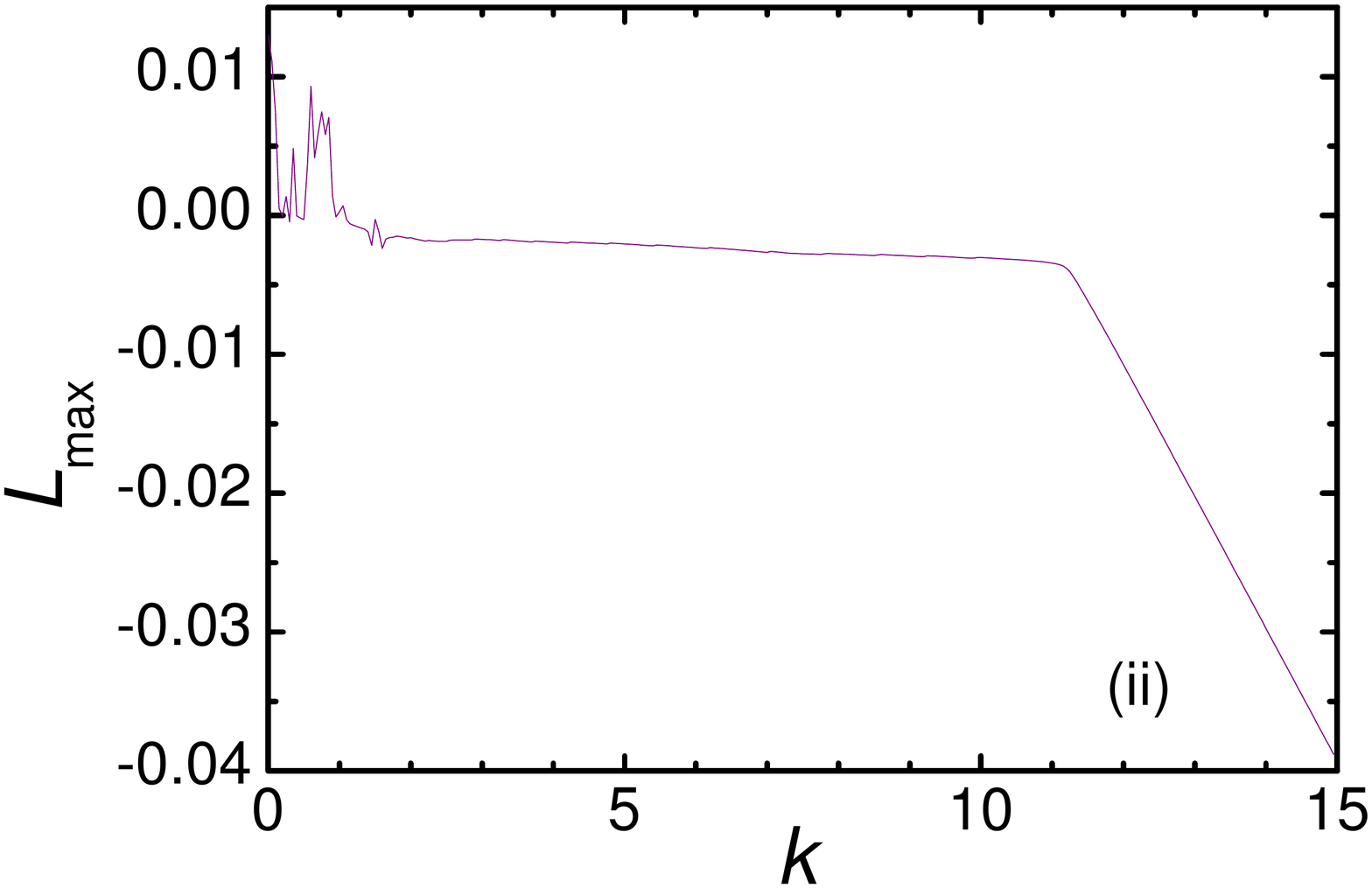}\\ \includegraphics[width=8cm, height=5cm]{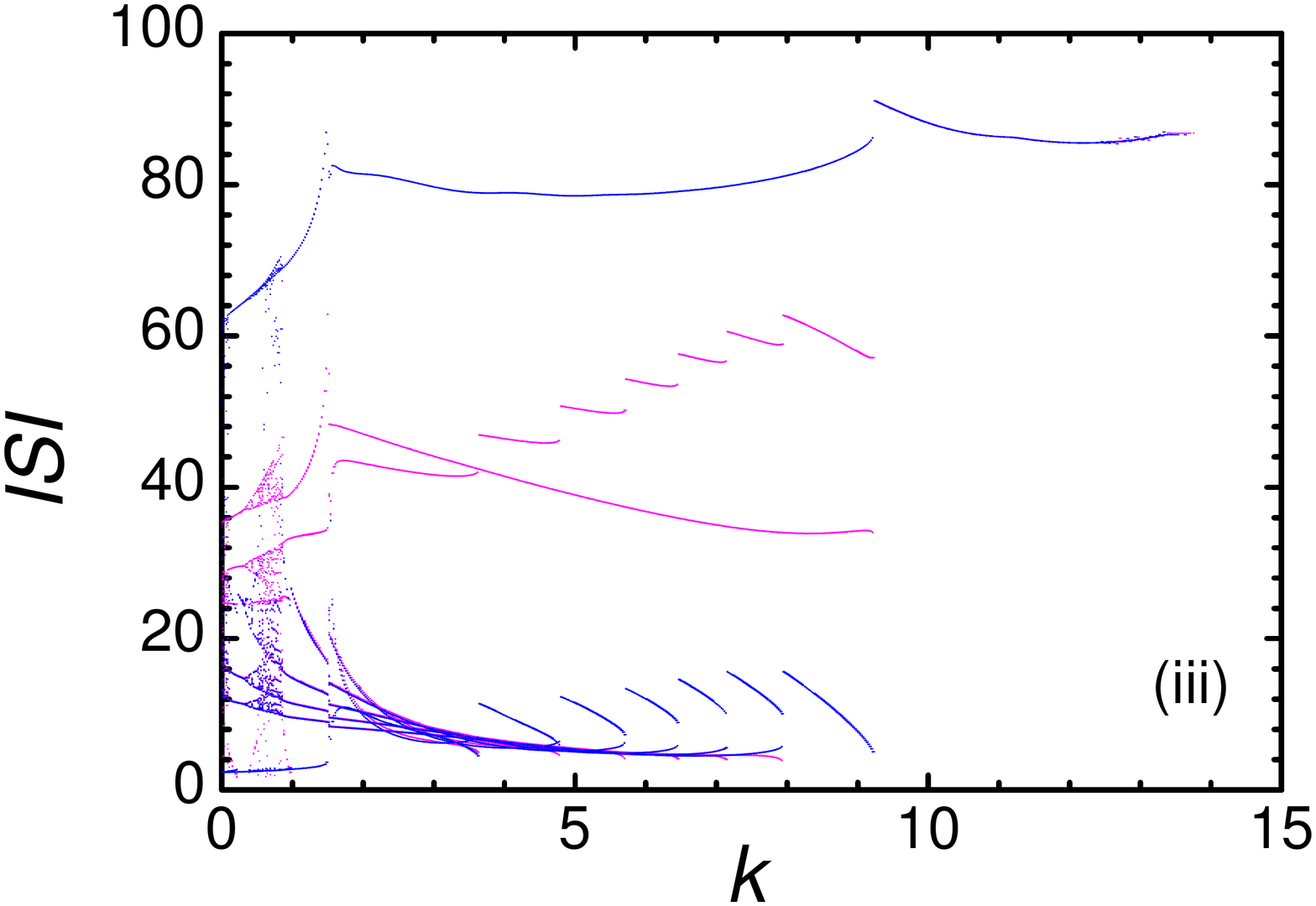} 
\caption{Bifurcation diagrams and  the largest
Lyapunov exponent as a function of $k$, for $r=0.008$, with $s=4$, $I_{ext}= 3.25$. 
(i) In these and following bifurcation diagrams, magenta and blue  lines correspond respectively to the local  minima and maxima of the timeseries $y(t)$, (ii) largest Lyapunov exponent.
 (iii) interspike intervals: time elapsed between maxima (blue) and between minima (magenta). After approx. $k>9$, the single value corresponds to the period of simple oscillations, which become damped near $k>11$. 
}
\label{fig:bif-k}
\end{figure}
 
Further details on the effects of varying $k$ are presented through the bifurcation diagrams  for the recovery current. 
In Fig.~\ref{fig:bif-k}, we use $k$ as bifurcation parameter when  $r=0.008$.
Besides the diagram for the extreme values  in Fig.~\ref{fig:bif-k}(i), we also show the corresponding plot of 
the largest Lyapunov exponent $L_{max}$ in Fig.~\ref{fig:bif-k}(ii). 
For $k>11$, the dynamics tends to a fixed point, characterized by a negative exponent (dark blue region in the diagrams of Fig.~\ref{fig:kr}). 
We also present the diagram for interspike intervals, $ISI$ as  function of $k$, in Fig.~\ref{fig:bif-k}(iii). The time intervals between consecutive maxima (minima) of $y(t)$ are ploted in blue (magenta).  The larger values between maxima correspond to the quiescent intervals, while the smaller ones are intraburst interspike intervals. For small $k$ the chaotic windows are also reflected in ISI. 
A single value means simple periodic oscillations as those illustrated in Fig.~\ref{fig:attractor}(c), that occur approximately within $9<k<11$, above that interval, oscillations are damped (tending to a constant value) but still detected by our code until the amplitude is so small that the machine precision is attained. 
As $k$ increases, the chaotic windows disappear, and $L_{max}$ becomes  negative for $k\gtrsim 2$.
Moreover, the characteristic time between spikes changes with $k$. At  
At $k\simeq 9$, simple periodic oscillations (without multiple spikes) occur, and at 
$k\simeq 11$ oscillations are lost, which means drastic consequences on neuron performance.
Cuts for other values of $r$  (0.001 and 0.05) are presented in Appendix~\ref{appd}, complementing the information of the heat-plots of Fig.~\ref{fig:kr}.

In the following figures ~\ref{fig:bifurcation-r}-\ref{fig:bifurcation-I}, we compare the bifurcation diagrams in the absence ($k=0$) and presence ($k=5$) of electromagnetic induction, using $r$, $s$ and $I_{ext}$ as bifurcation parameters, respectively.    
The local maxima (blue) and minima (magenta) of the timeseries are distinguished.  
The corresponding plots for the largest Lyapunov exponent $L_{max}$
are  also shown.

\begin{figure}[h!]
\centering
\includegraphics[width=8cm, height=5cm]{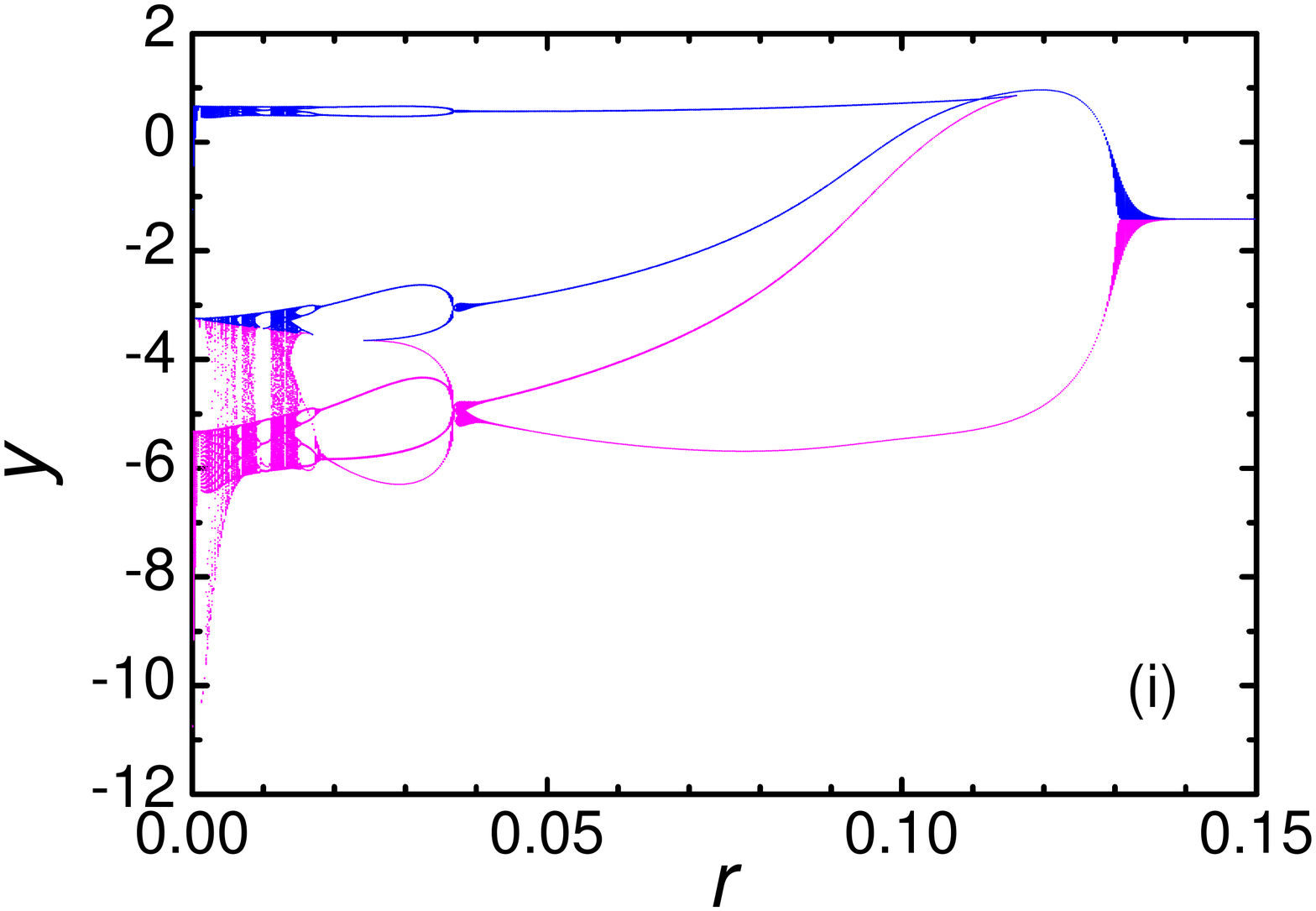}
\includegraphics[width=8cm, height=5cm]{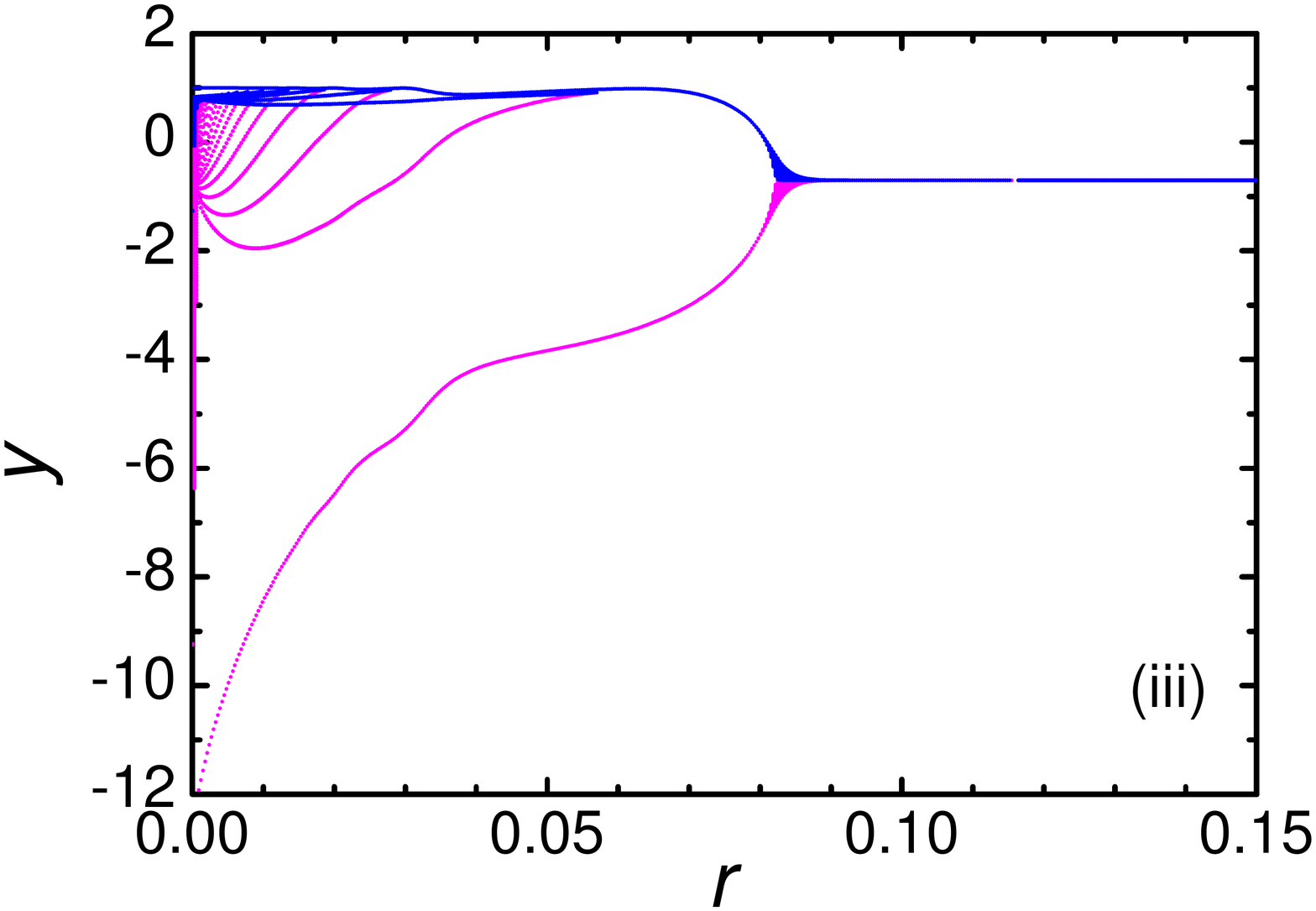}\\
\includegraphics[width=8cm, height=5cm]{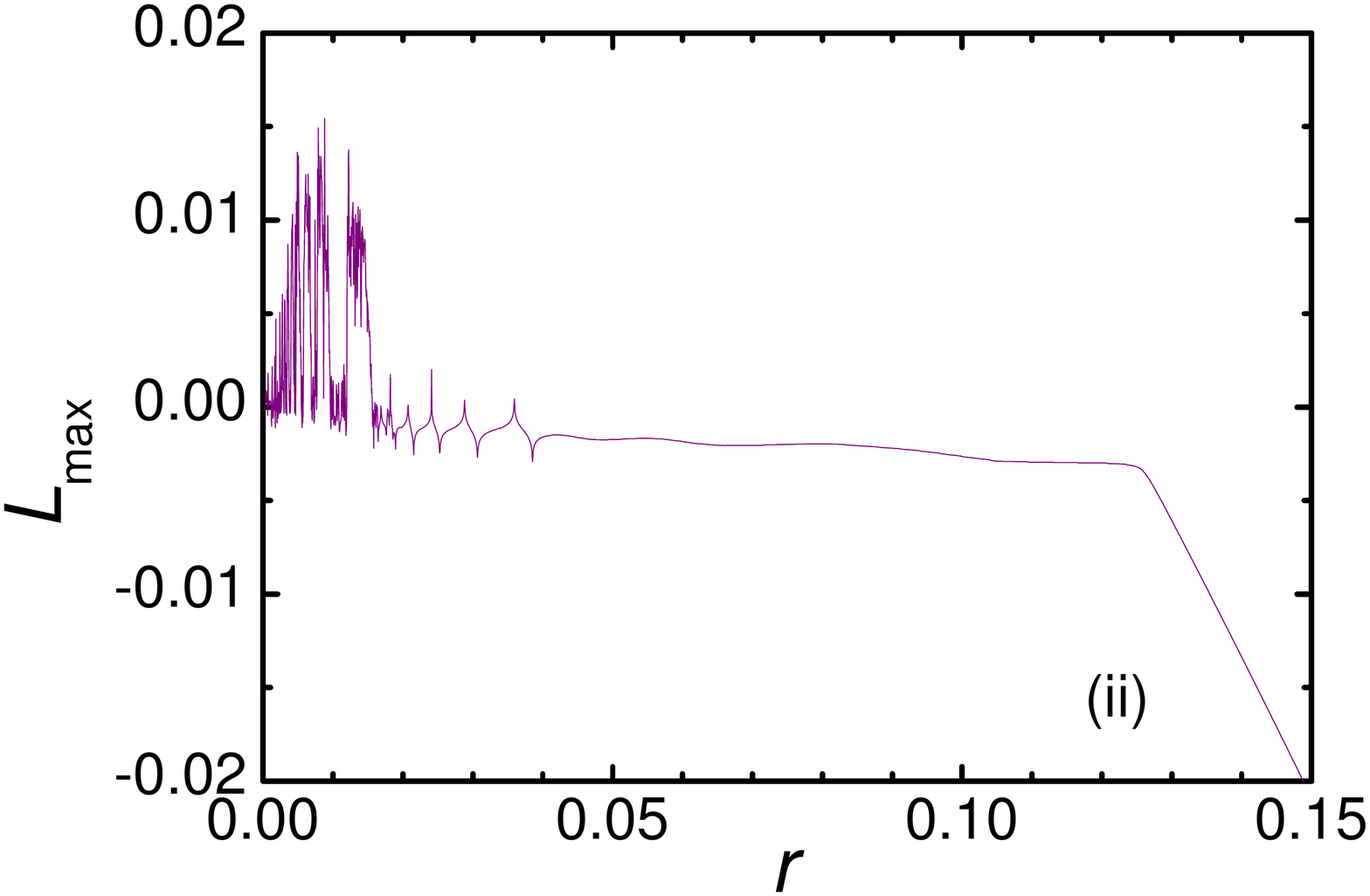} 
\includegraphics[width=8cm, height=5cm]{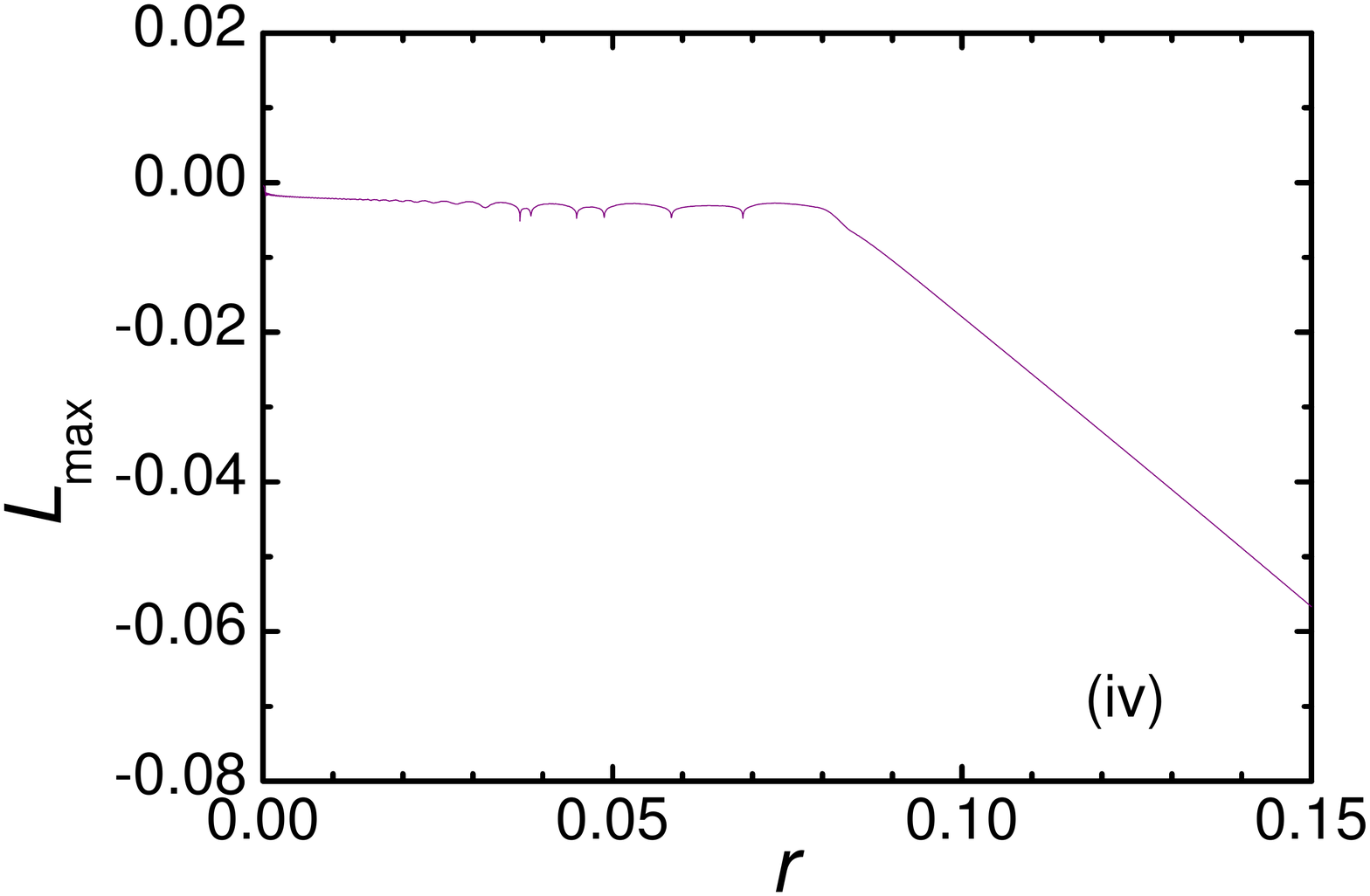}
\caption{Bifurcation diagrams and  the largest
Lyapunov exponent as a function of $r$, for $k=0$ (i)-(ii) and $k=5$ (iii)-(iv),  with $s=4$, $I_{ext}= 3.25$. 
In these and following bifurcation diagrams, magenta and blue  lines correspond respectively to the local  minima and maxima of the timeseries $y(t)$.
}
\label{fig:bifurcation-r}
\end{figure}

For the 
bifurcation parameter  $r$,  which rules 
the slow dynamics,   we show Fig.~\ref{fig:bifurcation-r}. 
In the case $k = 0$, corresponding to
Figs.~\ref{fig:bifurcation-r}.i and 
~\ref{fig:bifurcation-r}.ii,   we observe chaotic intervals, interrupted by regularity windows. 
In particular, a reverse period-doubling transition occurs.  
Therefore, decreasing the value of parameter $r$ can induce a complex sequence of period-doubling bifurcations, resulting in the duplication of spikes.  
When the electromagnetic radiation is taken into account, with intensity $k=5$, as shown in Figs.~\ref{fig:bifurcation-r}.iii and 
~\ref{fig:bifurcation-r}.iv, 
 the neuronal model no longer exhibits complex behaviors, and we find only periodic oscillations.
The diagrams for intermediate values of $k$, showing how increasing $k$ extinguishes the chaotic windows, are presented in the Appendix~\ref{appd}.
  
\begin{figure}[h!]
	\centering
\includegraphics[width=8cm, height=5cm]{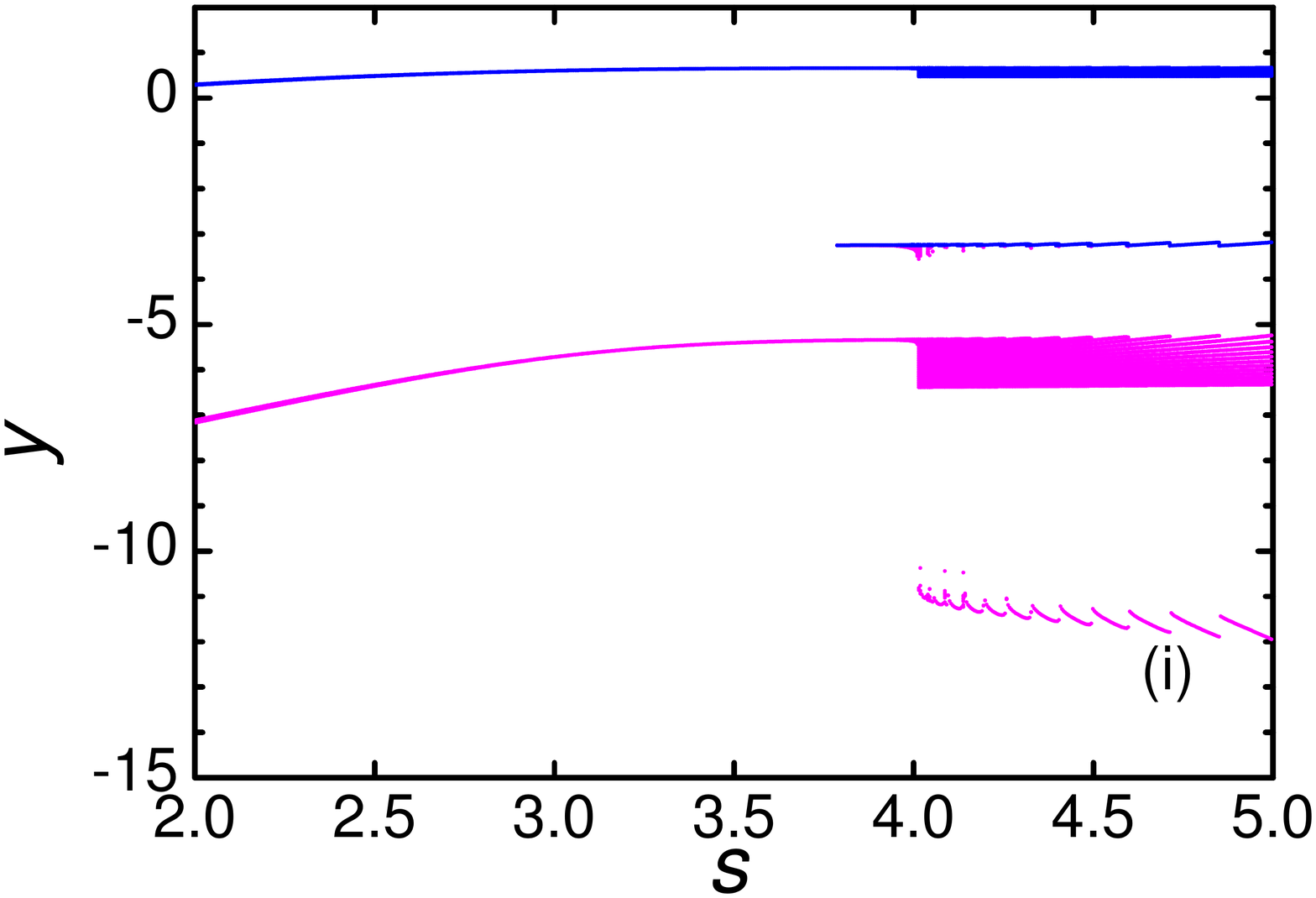}
\includegraphics[width=8cm, height=5cm]{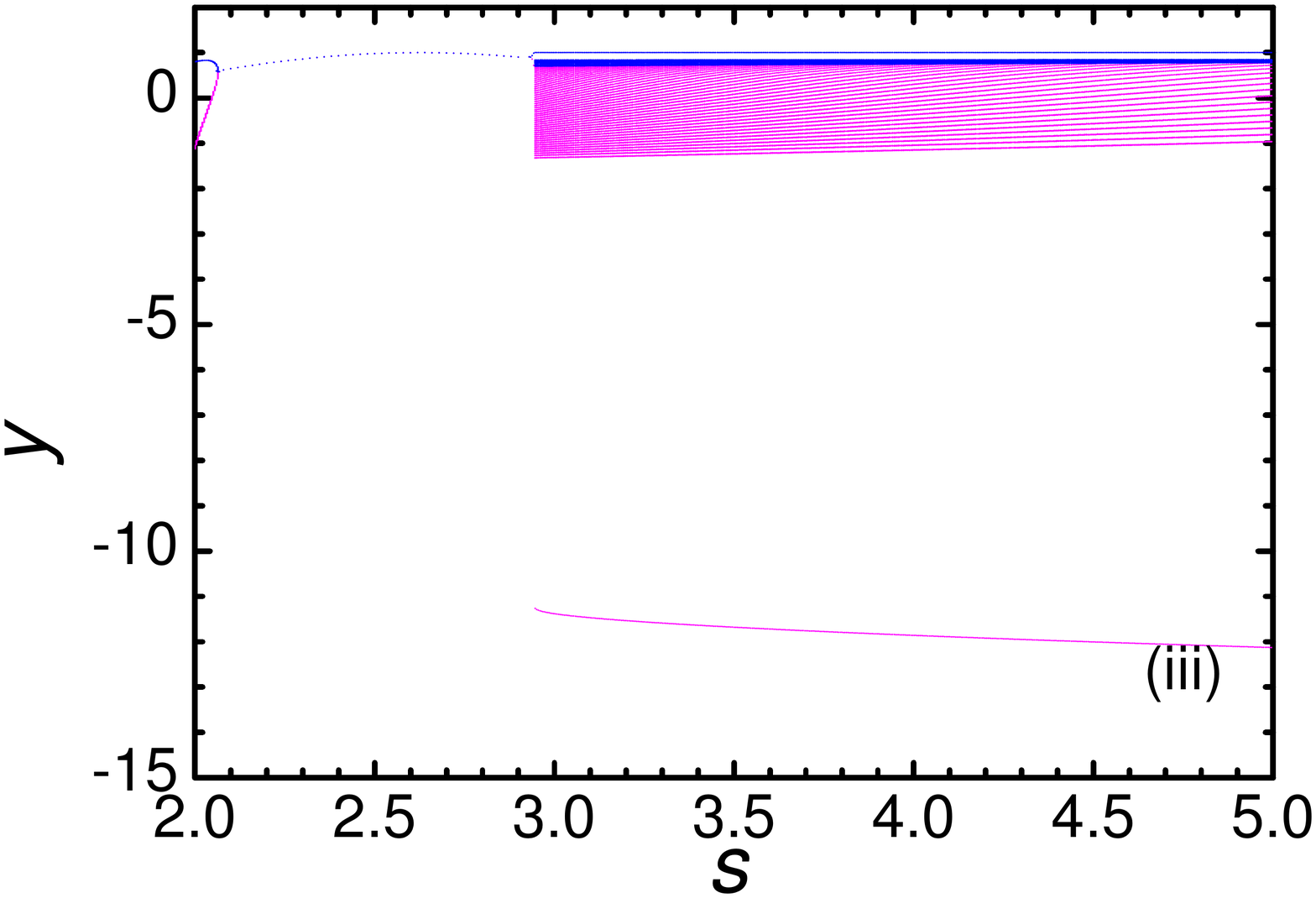}\\
\includegraphics[width=8cm, height=5cm]{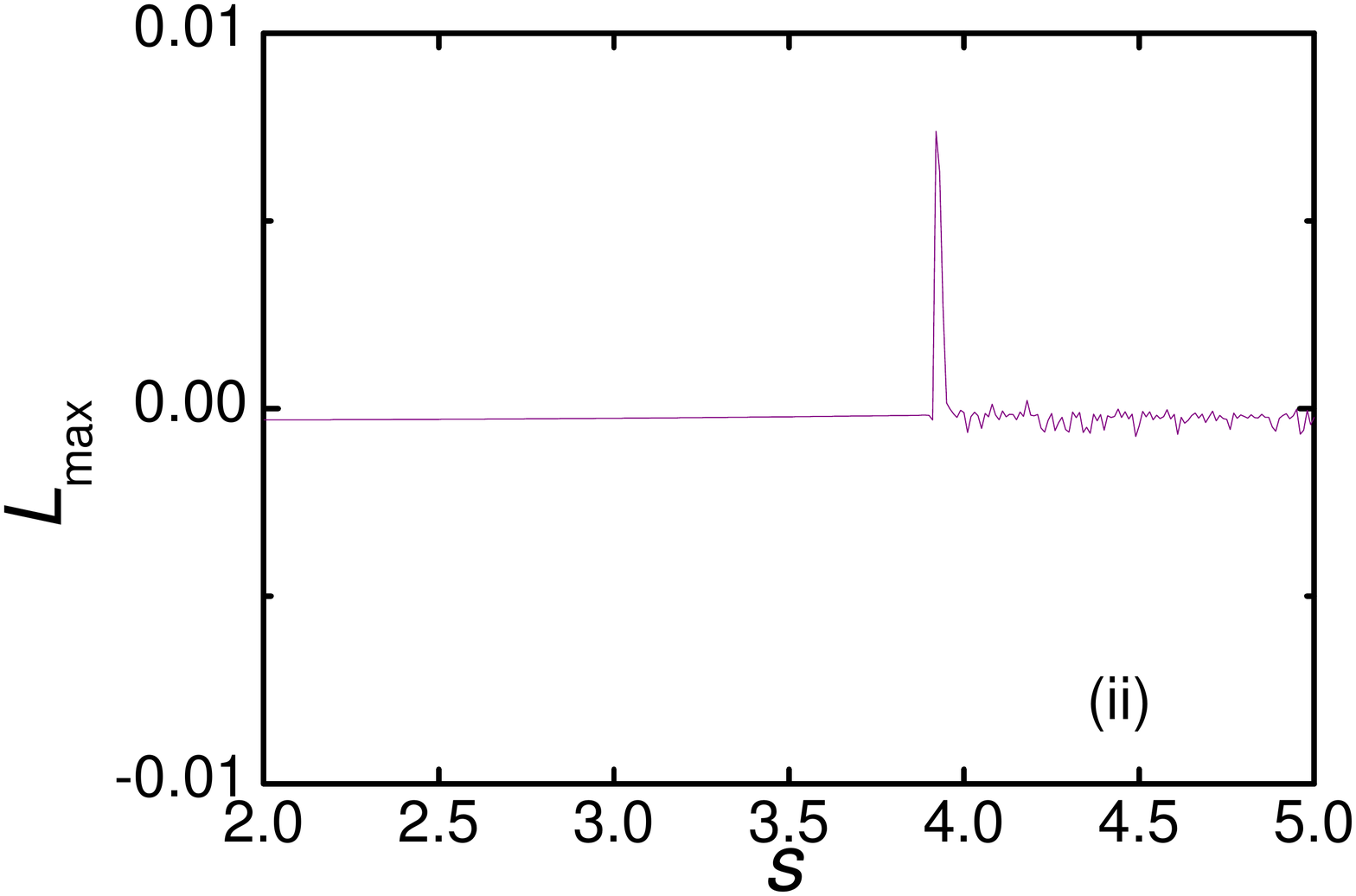}
\includegraphics[width=8cm, height=5cm]{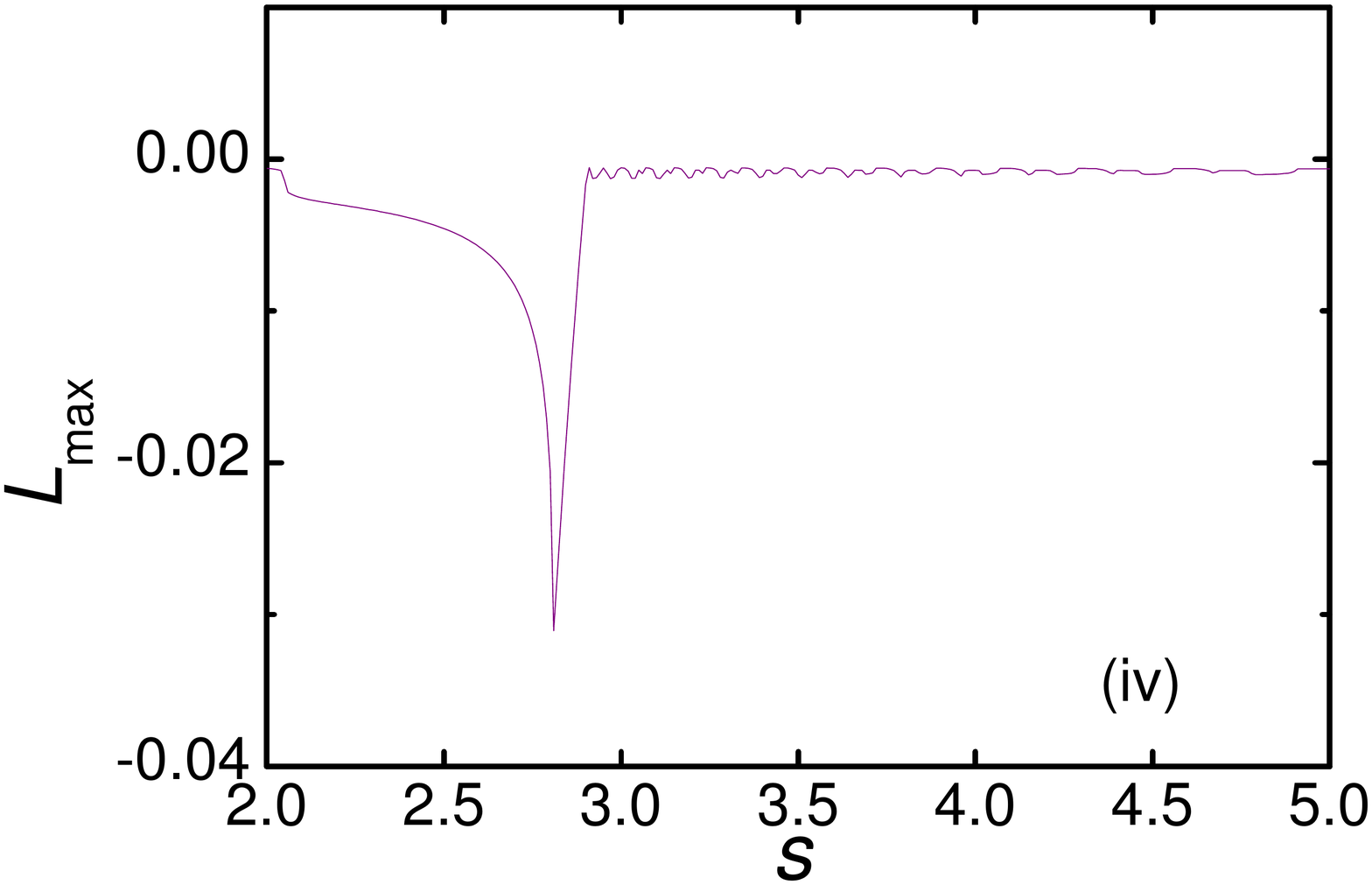}
	\caption{Bifurcation diagrams and  the largest
Lyapunov exponent as a function of $s$, for $k=0$ (i)-(ii) and $k=5$ (iii)-(iv),  with $r=0.001$, $I_{ext}= 3.25$. (For $s<3.0$ in (iii), a transient longer than $10^4$ was discarded, indicating a slow relaxation.)  
}
	\label{fig:bifurcation-s}
\end{figure}
Bifurcation diagrams are also shown  for the control parameters $s$ ( Fig.~\ref{fig:bifurcation-s}) and   $I_{ext}$ (Fig.~\ref{fig:bifurcation-I}). 
As a function of $s$,  oscillations with several spikes appear at $s\approx 4$ when $k=0$.  
The peak of $L_{max}$ at $s\simeq 3.9$ is a reflection of the transition between two different oscillatory regimes.  
For $k=5$, the oscillations occur for smaller $s\simeq 2.9$ but the amplitude of the spikes is diminished.
This reduction of spike amplitude  is also observed as a function of $I_{ext}$, in Fig.~\ref{fig:bifurcation-I}.

\begin{figure}[h!]
	\centering
\includegraphics[width=8cm, height=5cm]{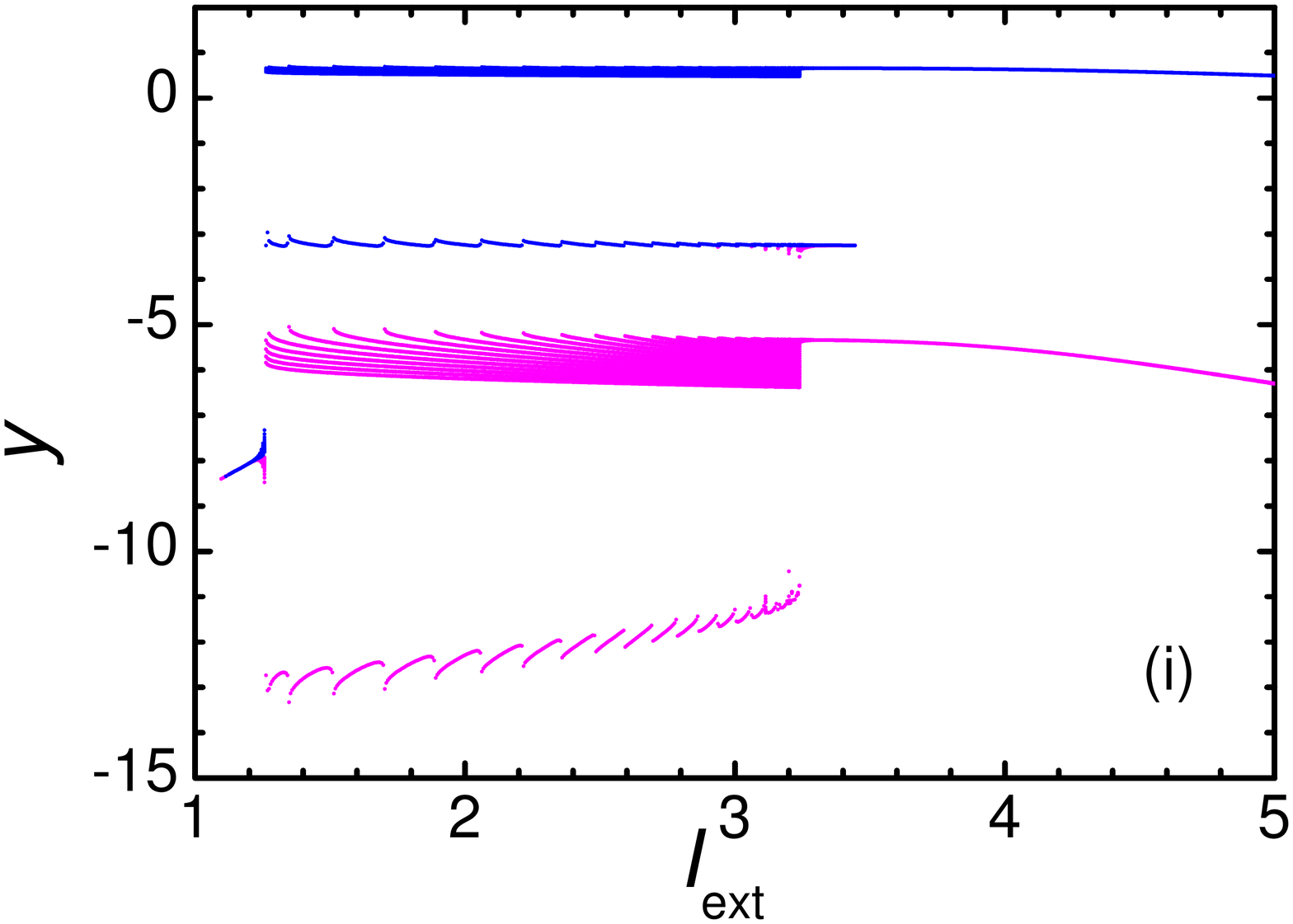}
\includegraphics[width=8cm, height=5cm]{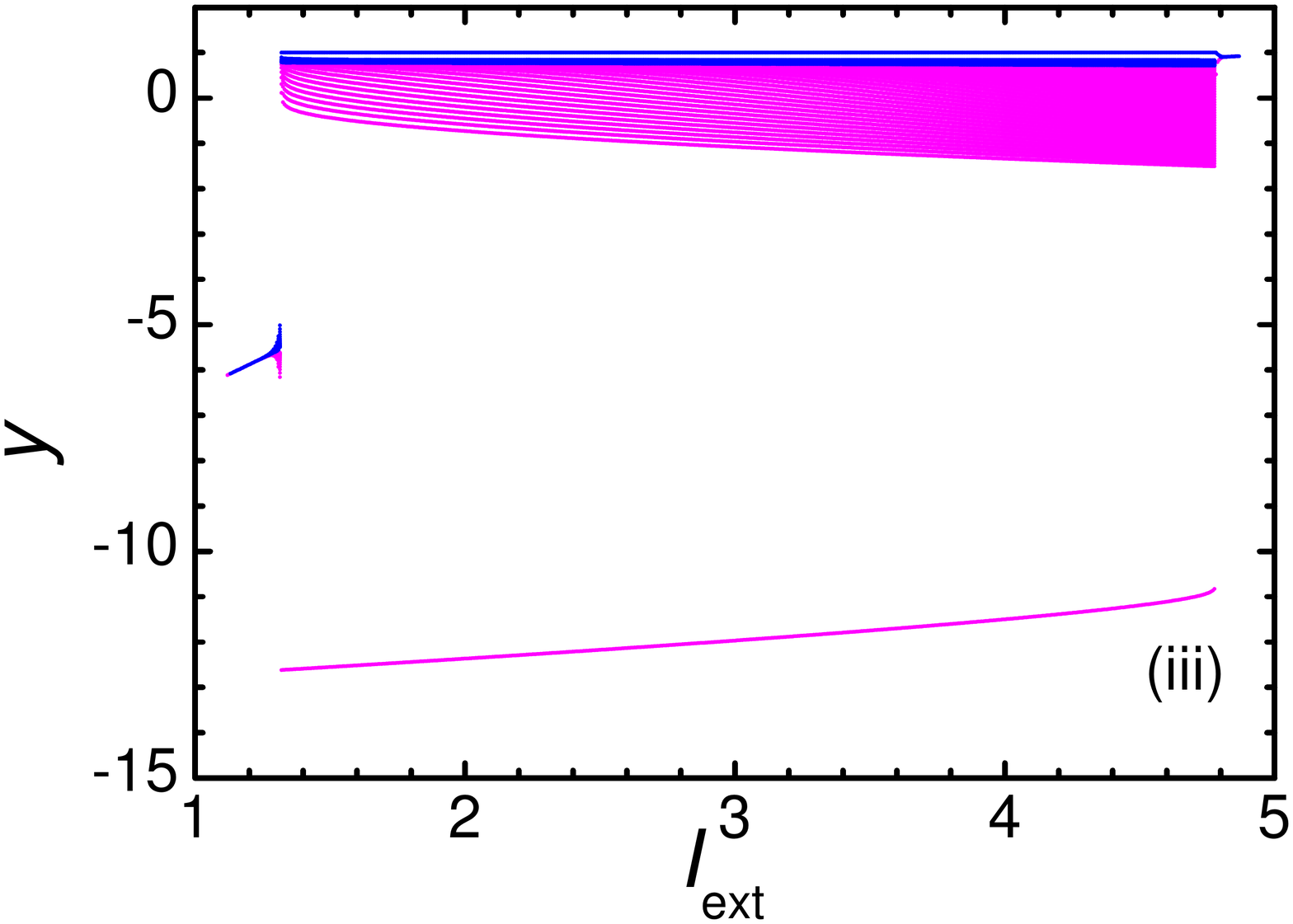}\\
\includegraphics[width=8.1cm, height=5cm]{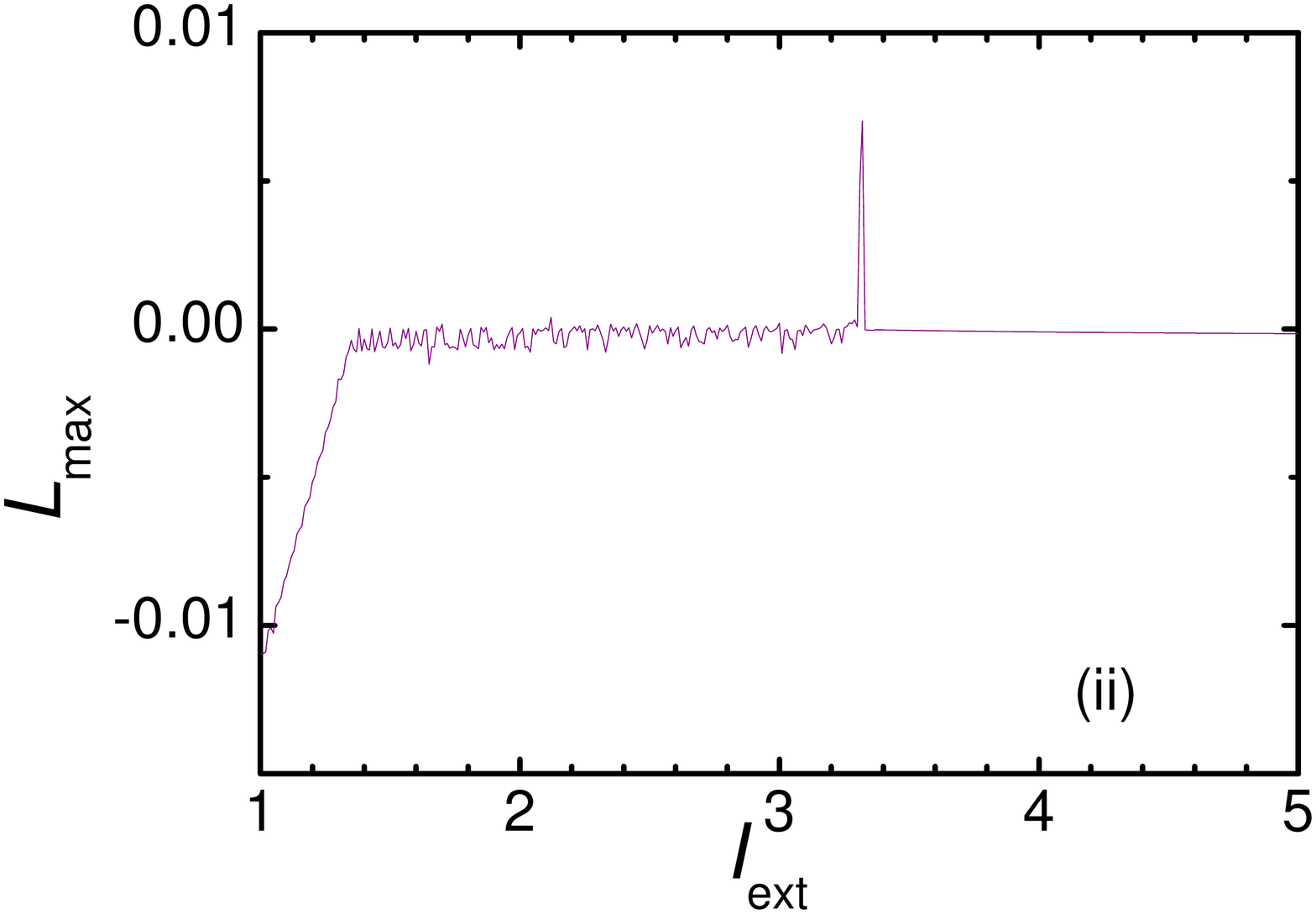}
\includegraphics[width=8.1cm, height=5cm]{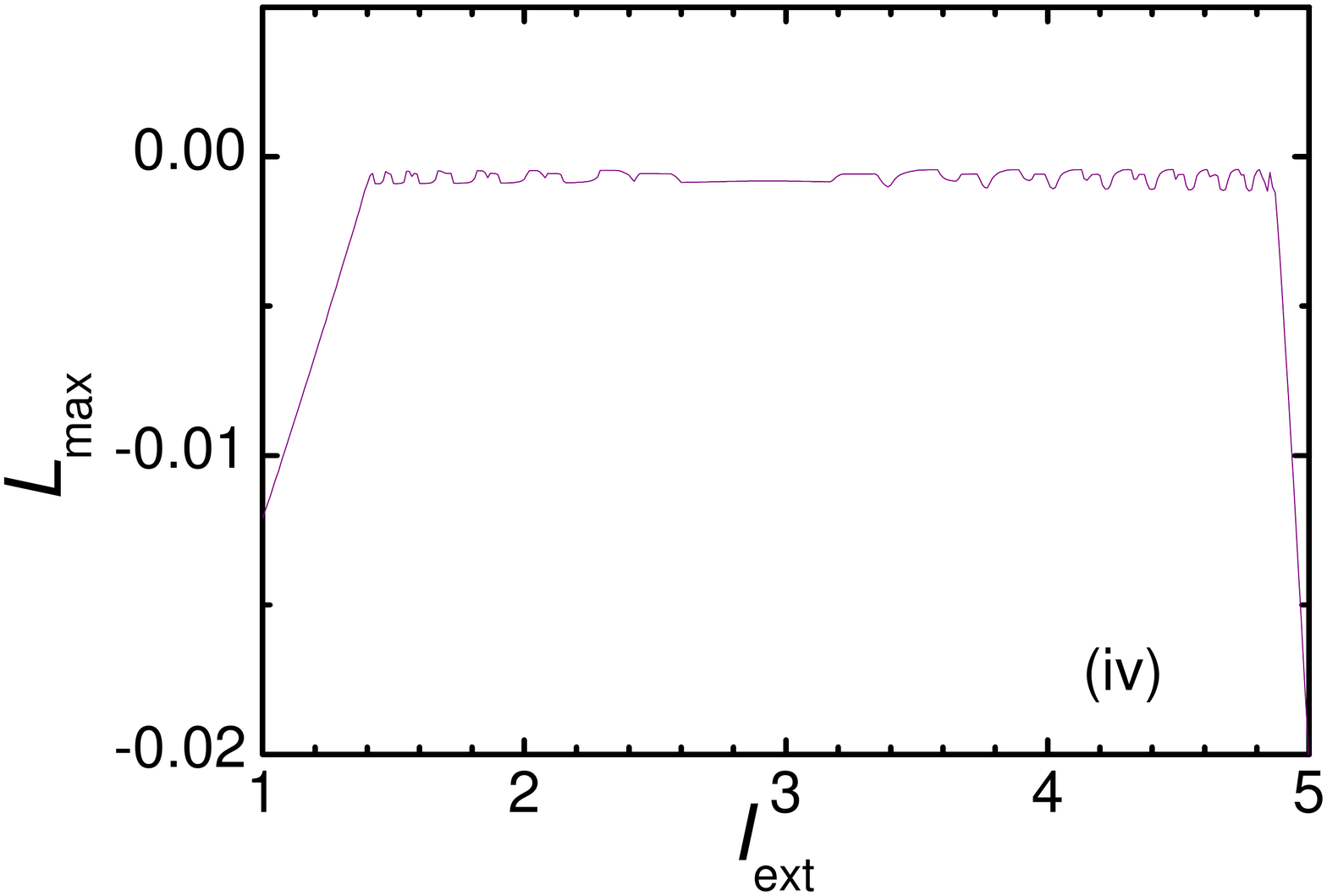}
	\caption{Bifurcation diagrams and  the largest
Lyapunov exponent as a function of $I=I_{ext}$, for $k=0$ (i)-(ii) and $k=5$ (iii)-(iv),  with
 $r=0.001$, $s=4$. 
}
	\label{fig:bifurcation-I}
\end{figure}
 
Therefore,  
  when electromagnetic induction is present, we observe the tendency that the chaotic windows 
disappears and the neuron exhibits regular spikes.

\section{Conclusions}
\label{sec:conclusions}

We have performed a stability analysis  of the Hindmarsh-Rose neuronal, extended to take into account the effect of the magnetic flux on the membrane potential (MHR model)~\cite{lv16}.
In Section~\ref{sec:dynamics}, we have shown the effects of electromagnetic induction 
on neuronal dynamics by varying the magnetic coupling $k$. We noted that  the domain of existence of three equilibrium points in the plane $s-I_{ext}$  decreased drastically when increasing $k$,
leading to a dynamics with a single equilibrium point. Moreover, unstable points become progressively stable when $k$ increases, spoiling oscillations. 
The observed bifurcations in the neuronal dynamics  reveal a
complex structure when the parameters $(I_{ext}, r, s)$ are changed in the absence of magnetic induction.
Variations of the maximal Lyapunov exponent, bifurcations diagrams, phase portraits, and time series allowed to emphasize the stabilizing and regularizing role of the introduction of electromagnetic induction on the neuron dynamics, for the values of the parameters considered.

Our results can be relevant  as basis for future studies on networks of MHR neurons.
As possible extensions, it would be also interesting to analyze the effect of different kinds of additive and multiplicative noises, as well as time-delays in the responses of the neuron.

  \section*{Acknowledgments}
C.A. acknowledges partial financial support from Brazilian agencies CAPES, CNPq and Faperj. 

\section*{Data availability statements}
The data from simulations that support the findings of this study are available on request from
the corresponding author, RY.

\appendix
\section{Equilibrium points}
\label{appa} 

With the change of variables $x_e = t-a_1/(3a_0)$,  Eq.~(\ref{eq:xe}) becomes reduced to $t^3+pt+q=0$, where
$p=(3a_0a_2-a_1^2)/(3a_0^2)$ and 
$q=(2a_1^3-9a_0\,a_1\,a_2+27a_0^2a_3)/(27a_0^3)$. 
To obtain its roots, we define the discriminant $\Delta = 
q^{2} + \frac{4}{27}p^{3}$, 
which, after substitution of 
the coefficients defined in Eq.~(\ref{eq:as}),  explicitly becomes
\begin{eqnarray}\nonumber
\Delta = 
\left[ \dfrac{2(b-d)^3-9T(b-d)(s+k\alpha)}{27T^3}+\frac{c+sx_0 +I_{ext}}{T}  \right]^2
+\frac{4}{27}\left[\dfrac{3T(s+k\alpha)-(b-d)^2}{3T^2} \right]^3\,.\;\;\;\;\;\\
\label{eq:discriminant} 
\end{eqnarray}
The signal of $\Delta$ determines the number of real roots. 
Then, setting $\Delta=0$, we extracted the expression $I_{ext}^\pm(s)$  in  Eq.~(\ref{eq:Ipm}), that delimits the regions with three (A) and one (B) real-valued  roots.

The real-valued solutions $x_e$ of Eq.~(\ref{eq:xe}) yield the equilibrium points of the form

\begin{equation}
    E=(x_{e},y_{e},z_{e},\phi_{e})= (x_{e},-d x_{e}^2+c,s(x_e-x_0), k_1x_{e}/k_2)\,.
\end{equation}

\begin{enumerate} 

\item If $\Delta$ < 0, corresponding to region (A) in Fig.~\ref{fig:sI},	there are three   equilibrium  points $ E_{1}=(x_{e1},y_{e1},z_{e1},\phi_{e1})$, $ E_{2}=(x_{e2},y_{e2},z_{e2},\phi_{e2})$  and  $ E_{3}=(x_{e3},y_{e3},z_{e3},\phi_{e3})$, given by:
\begin{eqnarray} \nonumber
x_{ek}=2\sqrt{-\frac{p}{3}}\cos\left( \frac{1}{3}\arccos\left( \frac{-q}{\sqrt{-4p^3/27}}\right) -\frac{2\pi (k-1)}{3}\right)+\dfrac{b-d}{3T}\;\;\;\;\; \mbox{for $k=1,2,3$}\,.\\ \label{eq:E3} 
\end{eqnarray}

	\item If $\Delta$ > 0, corresponding to the region (B) in Fig.~\ref{fig:sI}, there is only one real root, then the system has a single equilibrium point $ E=(x_{e},y_{e},z_{e},\phi_{e}) $  defined by
\begin{equation}
\label{eq:E1}
x_{e}=\sqrt[3]{\dfrac{-q+\sqrt{\Delta}}{2}} + \sqrt[3]{\dfrac{-q-\sqrt{\Delta}}{2}} +\dfrac{b-d}{3T}. 
\end{equation}

\item If $\Delta$ = 0 (borderlines in Fig.~\ref{fig:sI}), there are two equilibrium points, $ E_{1}=(x_{e1},y_{e1},z_{e1},\phi_{e1})$ and $ E_{2}=(x_{e2},y_{e2},z_{e2},\phi_{e2})$, defined
	by
\begin{equation}
\label{eq:E2}
	x_{e1} = 2\sqrt[3]{\dfrac{-q}{2}} +\dfrac{b-d}{3T}, 
	\;\;\;\;\;\;			\mbox{and}  \;\;\;\;\;\;
			x_{e2} = -\sqrt[3]{\dfrac{-q}{2}} +\dfrac{b-d}{3T}.
\end{equation}

\end{enumerate}

\section{Coefficients of the characteristic polynomial}
\label{appb}

Here we give the explicit expressions of the coefficients $\delta_i$,  with $1\le i \le  4$, of the characteristic polynomial in Eq.~(\ref{eq:deltai}), associated to the Jacobian matrix (\ref{eq:J}):

 \begin{eqnarray} \nonumber
 \delta_{1}&=& 3\beta kk_{1}^{2}x_{e}^{2}/k_2^2 +3a x_{e}^{2} + {\alpha}k -2bx_{e} +k_{2} +r + 1,\\[3mm] \nonumber
 \delta_{2}&=&   
 3a k_{2} x_{e}^{2} 
 + 3a r x_{e}^{2}  +3a x_{e}^{2}    
+  {\alpha}kk_{2} + r({\alpha}k+s)   
   -2b r x_{e} \\ \nonumber
&+&   
   {\alpha}k 
   -2b k_{2} x_{e} 
   +2(d-b) x_{e} +k_{2} r 
    + k_{2} +  r     \\ \nonumber
   &+&
 3\beta kk_{1}^{2}x_{e}^{2}[ 3  k_{2}  
  +   r 
   +  1
  ]  /  k_{2}^{2},\\[3mm] \nonumber
  \delta_{3}&=&
   3ak_{2} rx_{e}^{2} 
  + 3ak_{2} x_{e}^{2} 
 + 3a rx_{e}^{2} 
 + \alpha k k_{2}r  
  -2bk_{2} rx_{e}  + \alpha kk_{2}\\ \nonumber
  &+& r(\alpha k +s) 
  +2(d-b)k_{2} x_{e}  +2(d-b) rx_{e} +k_{2} rs +k_{2} r     \\ \nonumber
 &+&
 3\beta k k_1^2 x_{e}^{2}\,[ 3 k_{2}r  +    3 k_{2}  + 
 r  ] / k_{2}^{2},\\[3mm] \nonumber
  \delta_{4}&=& r\,k_2\,[9\beta kk_{1}^{2}x_{e}^{2}/k_2^2 +3a x_{e}^{2} + \alpha k  + s   +2(d-b) x_{e}].
 \end{eqnarray}

\section{Nature of the equilibrium points}
\label{appc}

In the following tables, for chosen points $P$ in  the 
plane $s-I_{ext}$ (Fig.~\ref{fig:nature}), we present the associated equilibrium point(s) $E$, together with 
the corresponding eigenvalues of the Jacobian matrix, emphasizing (in the last column)  the nature of the equilibrium points, for $k=0$ (Table \ref{Tab1}) and $k=10$ (Table \ref{Tab2}).

\newpage

\begin{table}[h!]
	\begin{center}
	\caption{Case $k=0$. 
	Equilibrium points and corresponding eigenvalues.}
	\label{Tab1}
		\begin{scriptsize}
\begin{tabular}{|c|l|l|l|}
\hline  
$\;\;\;\;P=\left( {\begin{array}{*{20}{c}}
s\\
I_{ext}
\end{array}} \right)\;\;\;\;$ &  Equilibrium point
						$ E=(x_{e},y_{e},z_{e},\phi_{e}) $ & Eigenvalues of $ J(x_{e},y_{e},z_{e},\phi_{e}) $ & Nature of the equilibrium point \\
\hline \hline 
$  {P_1}=\left( {\begin{array}{*{20}{c}}
-2\\
1
\end{array}} \right)$
&  $ E_{1}=(1.53,-10.74,-6.30,0.31) $  &  $\begin{array}{*{20}{l}}
{{\lambda _1} = { 0.578  + 3.57i}}\\
{{\lambda _2} = { 0.578  - 3.57i}}\\
{{\lambda _3} = { - 0.5}}\\
{{\lambda _4} = { -8.4\times 10^{-4}}}
\end{array}$
 & saddle-focus  \\  
 \hline\hline
 $  {P_2}=\left( {\begin{array}{*{20}{c}}
 1.5\\
 1
 \end{array}} \right)$
 &  $ E_{2}=(-0.77,-2.01,1.26,-0.15) $  &  $\begin{array}{*{20}{l}}
 {{\lambda _1} = { - 7.63  }}\\
 {{\lambda _2} = { - 0.49}}\\
 {{\lambda _3} = { 0.16}}\\
 {{\lambda _4} = { 1.57\times 10^{-4}}}
 \end{array}$
  & saddle  point \\
\hline\hline
 	  &$ E_{3a}=(-2.57,-32.0,4.75,-0.51) $ &$\begin{array}{*{20}{l}}
		  {{\lambda _1} = {-35.9 }}\\
		  {{\lambda _2} = { -0.5}}\\
		  {{\lambda _3} = {-0.26}}\\
		  {{\lambda _4} = { -4.7\times 10^{-4}}}
		  \end{array}$& {\bf stable} node \\
		\cline{2-4}  $  {P_3}=\left( {\begin{array}{*{20}{c}}
		-5\\
	0
		\end{array}} \right) $ & $ E_{3b}=(2.19,-22.9,-19.0,0.44) $ & $\begin{array}{*{20}{l}}
		  {{\lambda _1} = {{-1.11} +4.67i}}\\
		  {{\lambda _2} = { {-1.11} - 4.67i}}\\
		  {{\lambda _3} = { -0.5}}\\
		  {{\lambda _4} = { -7.8\times 10^{-4}}}
		  \end{array}$& {\bf stable} saddle-focus   \\  
		\cline{2-4} &$ 
		E_{3c}=(-1.61,-12.1,0.0,-0.32) $ &$\begin{array}{*{20}{l}}
		  {{\lambda _1} = {-18.5}}\\
		  {{\lambda _2} = {-0.5}}\\
		  {{\lambda _3} = { 0.078}}\\
		  {{\lambda _4} = { 0.0025}}
		  \end{array}$ & saddle  point \\ 
	\hline\hline
		  &$ E_{4a}=(-2.41,-28.01,2.37,-0.48) $ &$\begin{array}{*{20}{l}}
		  {{\lambda _1} = {-32.62 }}\\
		  {{\lambda _2} = { -0.5}}\\
		  {{\lambda _3} = {-0.24}}\\
		  {{\lambda _4} = { -6.1\times 10^{-4}}}
		  \end{array}$& {\bf stable} node \\
		\cline{2-4}  $  {P_4}=\left( {\begin{array}{*{20}{c}}
		-3\\
		-1
		\end{array}} \right) $ & $ E_{4b}=(1.64,-12.43,-9.77,0.328) $ & $\begin{array}{*{20}{l}}
		  {{\lambda _1} = {{0.39} +3.80i}}\\
		  {{\lambda _2} = { {0.39} - 3.80i}}\\
		  {{\lambda _3} = { -0.5}}\\
		  {{\lambda _4} = { -7.9\times 10^{-4}}}
		  \end{array}$& saddle-focus   \\  
		\cline{2-4} &$ E_{4c}=(-1.23,-6.56,-1.16,-0.246) $ &$\begin{array}{*{20}{l}}
		  {{\lambda _1} = {-12.95}}\\
		  {{\lambda _2} = {-0.5}}\\
		  {{\lambda _3} = { 0.0359}}\\
		  {{\lambda _4} = { -0.00726}}
		  \end{array}$ & saddle  point \\
		\hline\hline
		 &$ E_{5a}=(1.54,-10.97,-9.49,0.309) $ &$\begin{array}{*{20}{l}}
		  {{\lambda _1} = {0.56+3.59i}}\\
		  {{\lambda _2} = {0.56-3.59i}}\\
		  {{\lambda _3} = {-0.5}}\\
		  {{\lambda _4} = { -7.74\times 10^{-4}}}
		  \end{array}$  &  saddle-focus \\ 
		\cline{2-4} $  {P_5}=\left( {\begin{array}{*{20}{c}}
		-3\\
		-2
		\end{array}} \right) $&  $ E_{5b}=(-2.58,-32.37,2.89,-0.516) $ & $\begin{array}{*{20}{l}}
		  {{\lambda _1} = {-36.1}}\\
		  {{\lambda _2} = {-0.5}}\\
		  {{\lambda _3} = { -0.266}}\\
		  {{\lambda _4} = { -6.88 \times 10^{-4}}}
		  \end{array}$ & {\bf stable} node \\
		\cline{2-4} &  $ E_{5c}=(-0.963,-3.64,-1.96,-0.192) $  & $\begin{array}{*{20}{l}}
		  {{\lambda _1} = {-9.67}}\\
		  {{\lambda _2} = {-0.5}}\\
		  {{\lambda _3} = { 0.11}}\\
		  {{\lambda _4} = { -3.7\times 10^{-3}}}
		  \end{array}$ &  saddle  point\\
		 \hline
		 \end{tabular}
		 	\end{scriptsize} 
		 	\end{center}
		 \end{table}

	\begin{table}[h!]
		\begin{center}
	\caption{Case $k=10$.
Equilibrium points and corresponding eigenvalues.}
		 \label{Tab2}
		\begin{scriptsize}
 \begin{tabular}{|c|l|l|l|}
\hline  
$P=\left( {\begin{array}{*{20}{c}}
s\\
I_{ext}
\end{array}} \right)$ &  Equilibrium point
						$ E=(x_{e},y_{e},z_{e},\phi_{e}) $ & Eigenvalues of $ J(x_{e},y_{e},z_{e},\phi_{e}) $ & Nature of the equilibrium point \\
\hline \hline  
 $  \;\;\;\;{P_1}=\left( {\begin{array}{*{20}{c}}
 -2\\
 1
 \end{array}} \right)\;\;\;\;$
 &  $ E_{1}=(1.379,-8.5,-5.99,0.275) $  &  $\begin{array}{*{20}{l}}
 {{\lambda _1} = { 0.21  +3.52i}}\\
 {{\lambda _2} = { 0.21  - 3.52i}}\\
 {{\lambda _3} = { - 0.505}}\\
 {{\lambda _4} = { -8.41\times 10^{-4}}}
 \end{array}$
  & saddle-focus  \\  
  \hline\hline
  $  {P_2}=\left( {\begin{array}{*{20}{c}}
  1.5\\
  1
  \end{array}} \right)$
  &  $ E_{2}=(-0.199,0.801,2.12,-0.039) $  &  $\begin{array}{*{20}{l}}
  {{\lambda _1} = { - 3.21}}\\
  {{\lambda _2} = { - 0.0989}}\\
  {{\lambda _3} = { - 0.0057}}\\
  {{\lambda _4} = { - 0.498}}
  \end{array}$
   & {\bf stable} node \\
   \hline\hline
   $  {P_3}=\left( {\begin{array}{*{20}{c}}
     -5\\
     0
     \end{array}} \right)$
     &  $ E_{3}=(1.79,-15.0,-17.0,0.36) $  &  $\begin{array}{*{20}{l}}
     {{\lambda _1} = { -0.54+4.23i}}\\
     {{\lambda _2} = { -0.54-4.23i }}\\
     {{\lambda _3} = { -0.51}}\\
     {{\lambda _4} = { - 7.3\times 10^{-4}}}
     \end{array}$
      &  {\bf stable} node-focus \\
 \hline	 \hline		
   $  {P_4}=\left( {\begin{array}{*{20}{c}}
     -3\\
     -1
     \end{array}} \right)$
     &  $ E_{4}=(1.32,-7.76,-8.83,0.265) $  &  $\begin{array}{*{20}{l}}
     {{\lambda _1} = { 0.282+3.426} i}\\
     {{\lambda _2} = { 0.282-3.426} i}\\
     {{\lambda _3} = { - 0.505}}\\
     {{\lambda _4} = { - 7.49\times 10^{-4}}}
     \end{array}$
      &   saddle-focus \\
 		\hline\hline
 		&$ E_{5a}=(1.37,-8.5,-8.9,0.27) $ &$\begin{array}{*{20}{l}}
 		  {{\lambda _1} = {0.229+3.51i}}\\
 		  {{\lambda _2} = {0.229-3.51i}}\\
 		  {{\lambda _3} = { -7.6\times 10^{-4}}}\\
 		  {{\lambda _4} = { - 0.00194}}
 		  \end{array}$  &   saddle-focus\\ 
 		\cline{2-4} $  {P_5}=\left( {\begin{array}{*{20}{c}}
 		-3\\
 		-2
 		\end{array}} \right) $&  $ E_{5b}=(-1.77,-14.78,0.477,-0.355) $ & $\begin{array}{*{20}{l}}
 		  {{\lambda _1} = {-22.7}}\\
 		  {{\lambda _2} = {-0.4832}}\\
 		  {{\lambda _3} = { -0.187}}\\
 		  {{\lambda _4} = { -2.48\times 10^{-4}}}
 		  \end{array}$ & {\bf stable}  node  \\
 		\cline{2-4} &  $ E_{5c}=(-1.46,-9.76,-0.451,-0.293) $  & $\begin{array}{*{20}{l}}
 		  {{\lambda _1} = {-17.3}}\\
 		  {{\lambda _2} = {-0.487}}\\
 		  {{\lambda _3} = { -0.122}}\\
 		  {{\lambda _4} = {4.5\times 10^{-4}}}
 		  \end{array}$ &  saddle  point\\
 		 \hline 
\end{tabular}
	\end{scriptsize} 
	\end{center}
\end{table}

\newpage
\section{Bifurcation diagrams}
\label{appd}

\renewcommand{\thefigure}{D\arabic{figure}}

\setcounter{figure}{0}

\begin{figure}[h!]
\centering
\includegraphics[width=7cm, height=4cm]{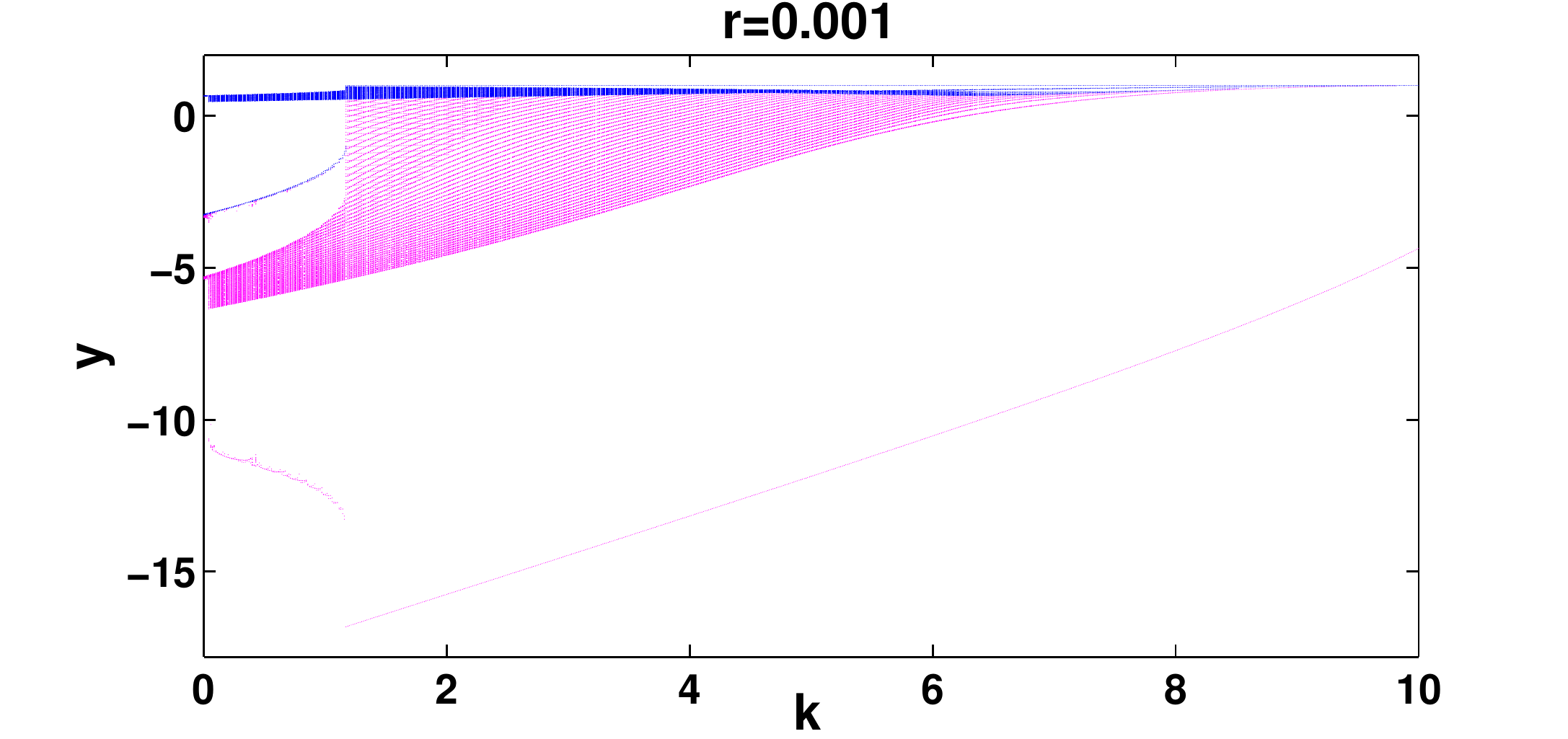}(i)
\includegraphics[width=7.cm, height=4cm]{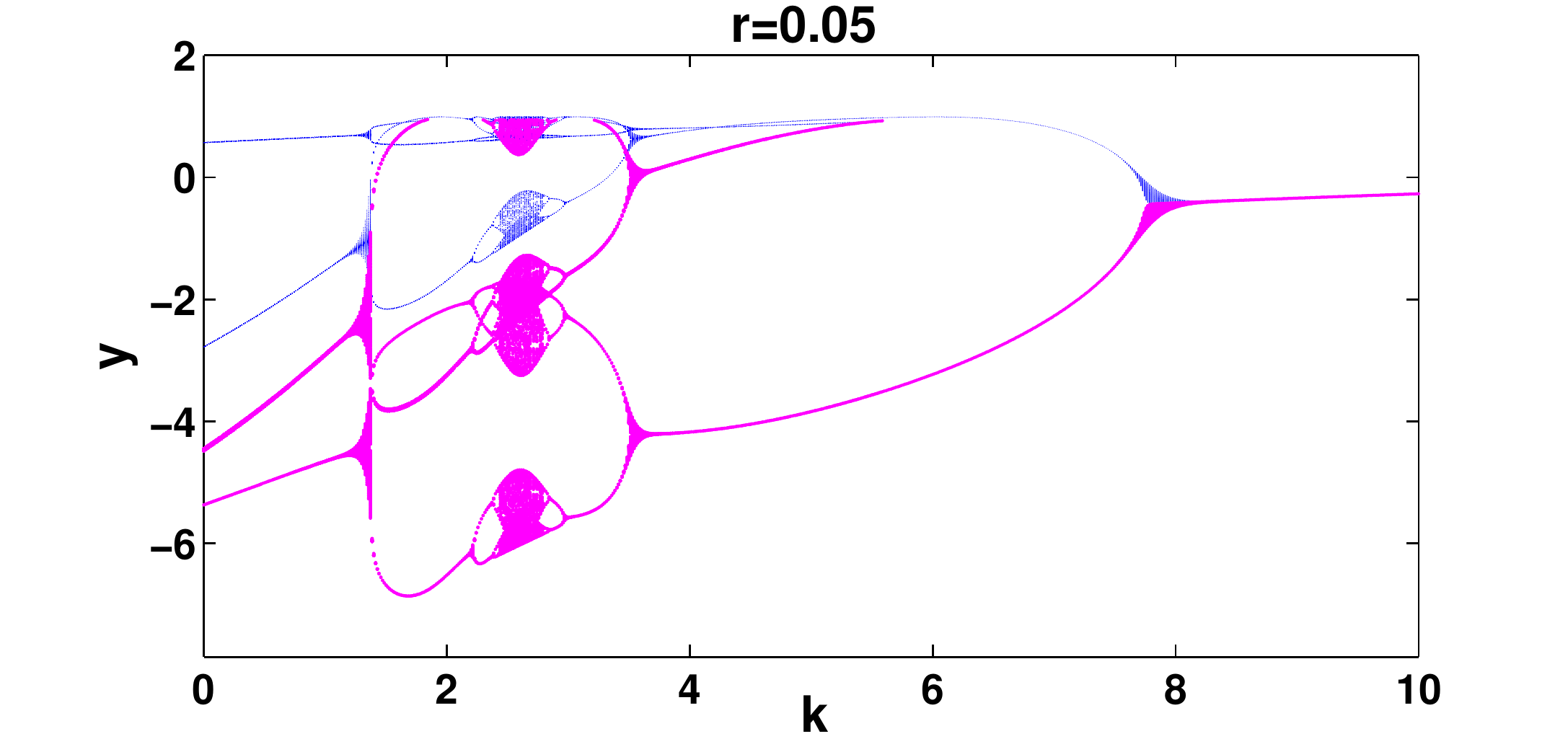}(iii)\\
\includegraphics[width=7.cm, height=4cm]{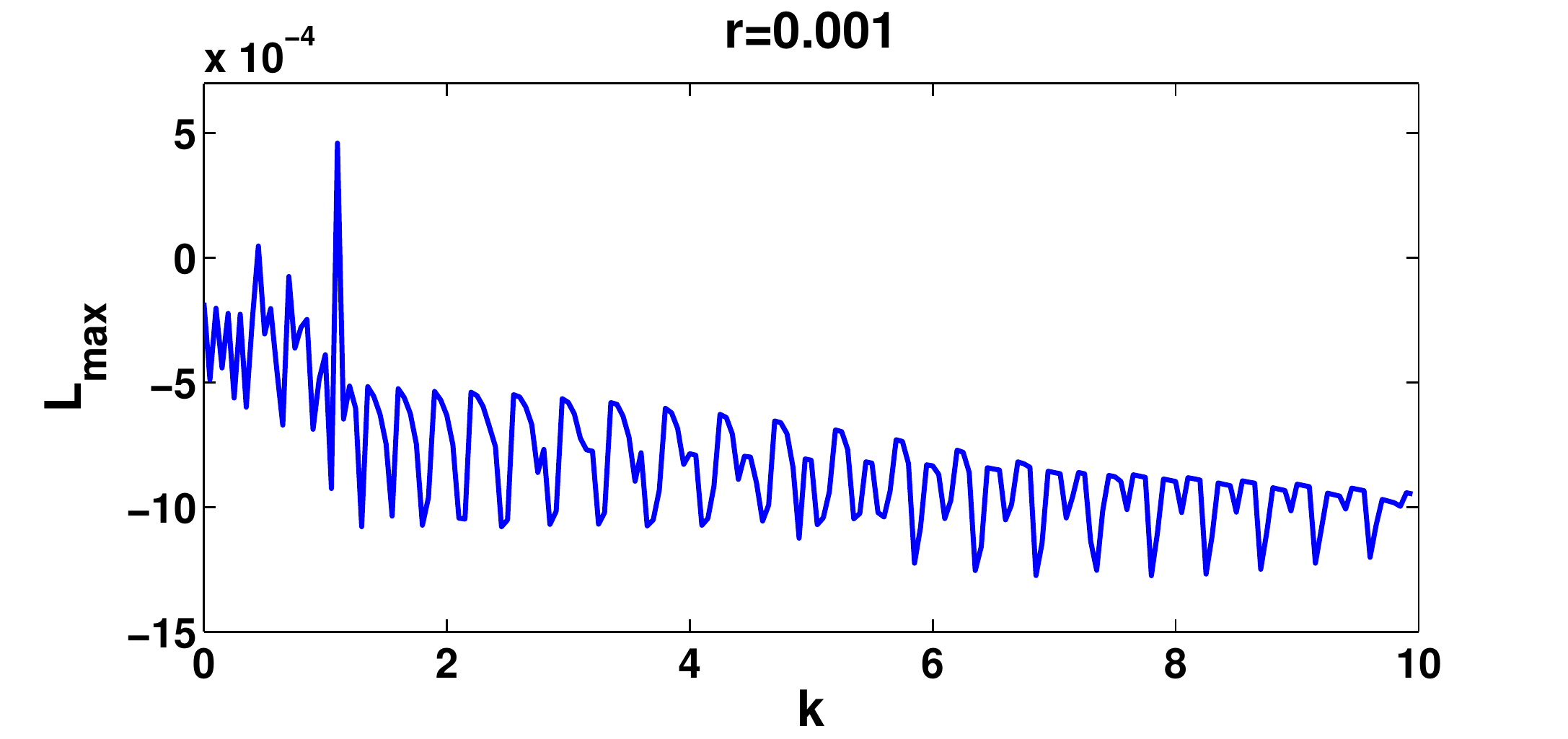}(ii)
\includegraphics[width=7.cm, height=4cm]{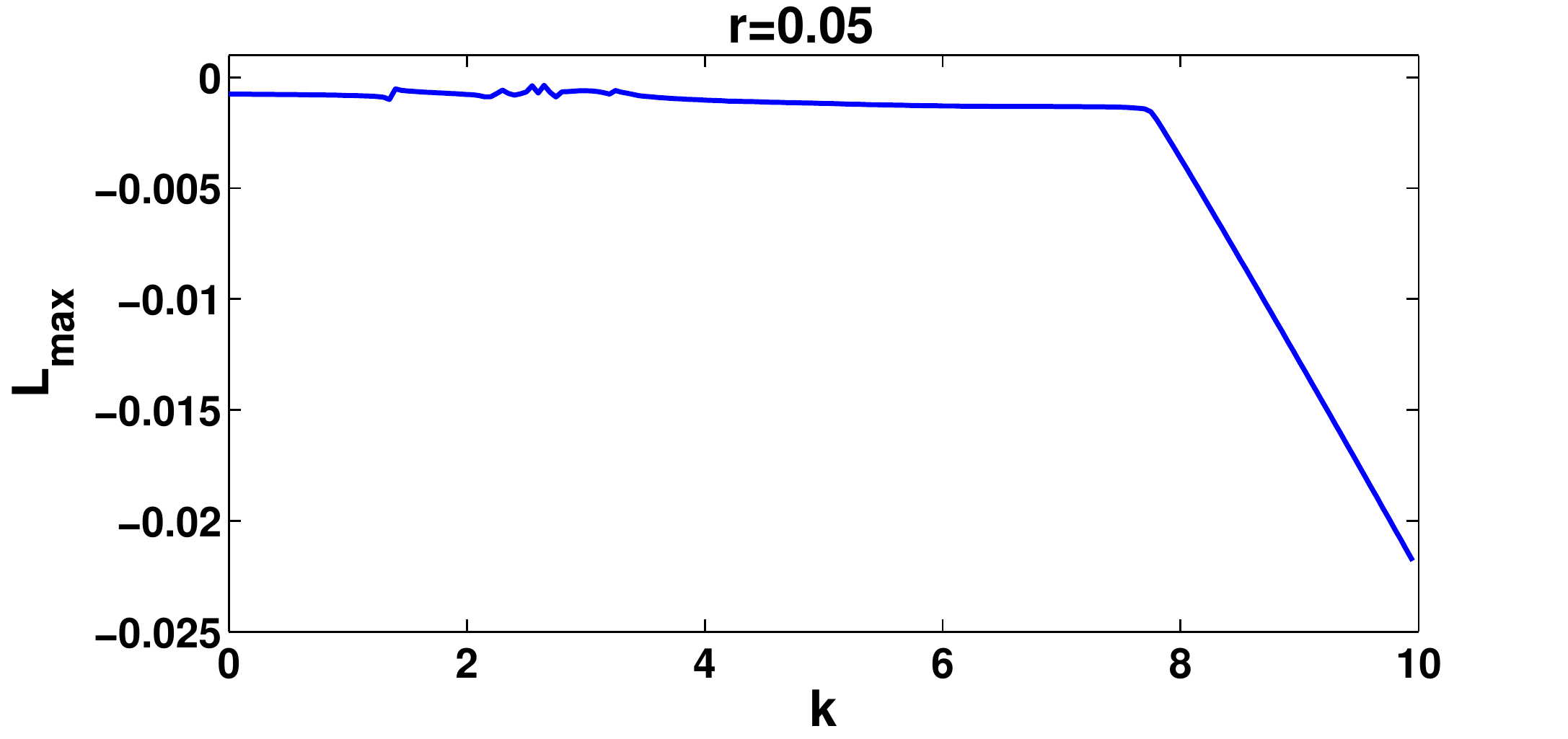}(iv)
\caption{Bifurcation diagrams and  the largest
Lyapunov exponent as a function of $k$, for $r=0.001$ (i)-(ii) and $r=0.05$ (iii)-(iv),  with $s=4$, $I_{ext}= 3.25$. 
Magenta and blue  lines correspond respectively to the local  minima and maxima of the timeseries $y(t)$.
}
\label{fig:bif-k-extra}
\end{figure}

\begin{figure}[h!]
	\centering
\includegraphics[width=7cm,height=4cm]{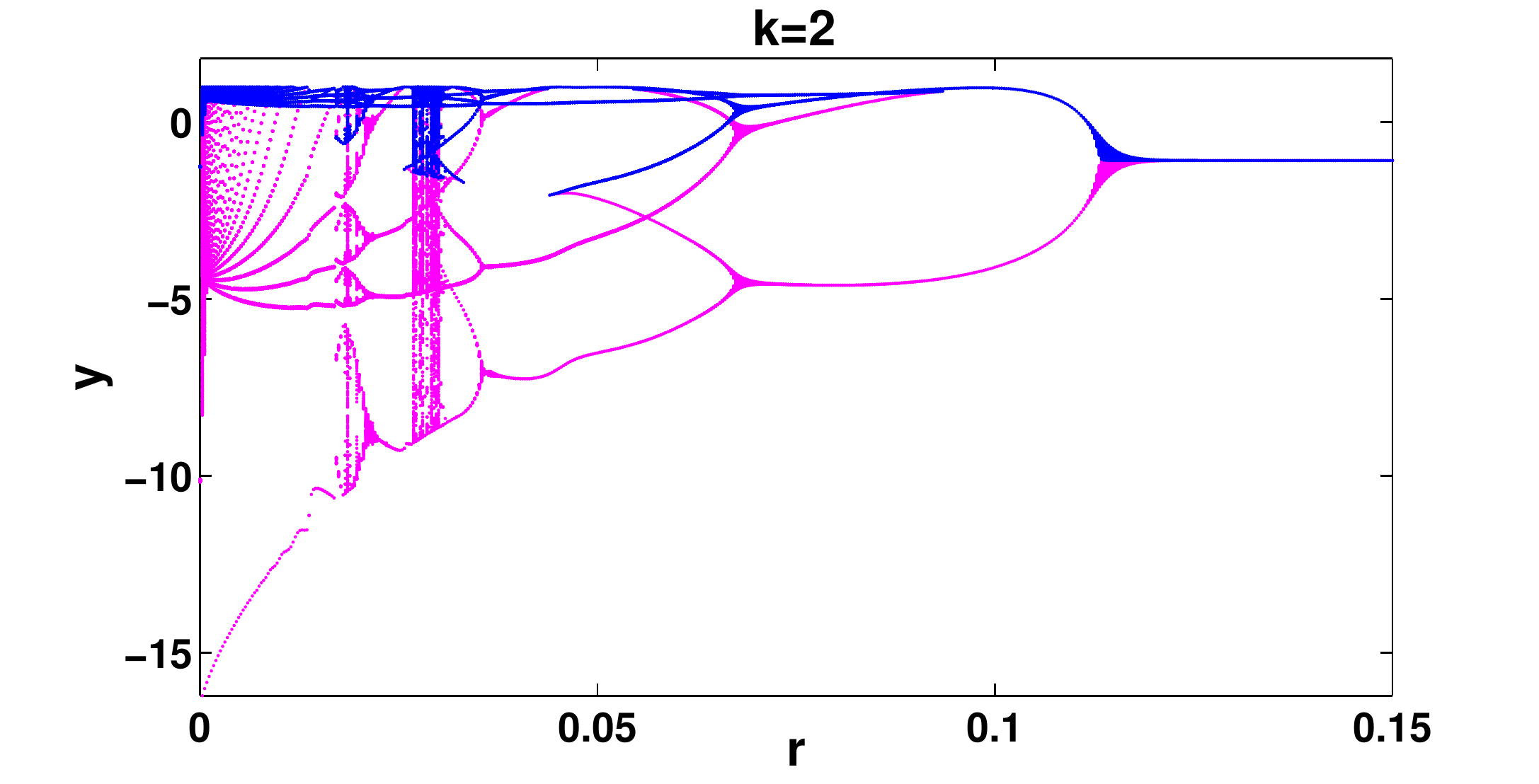}
\includegraphics[width=7cm,height=4cm]{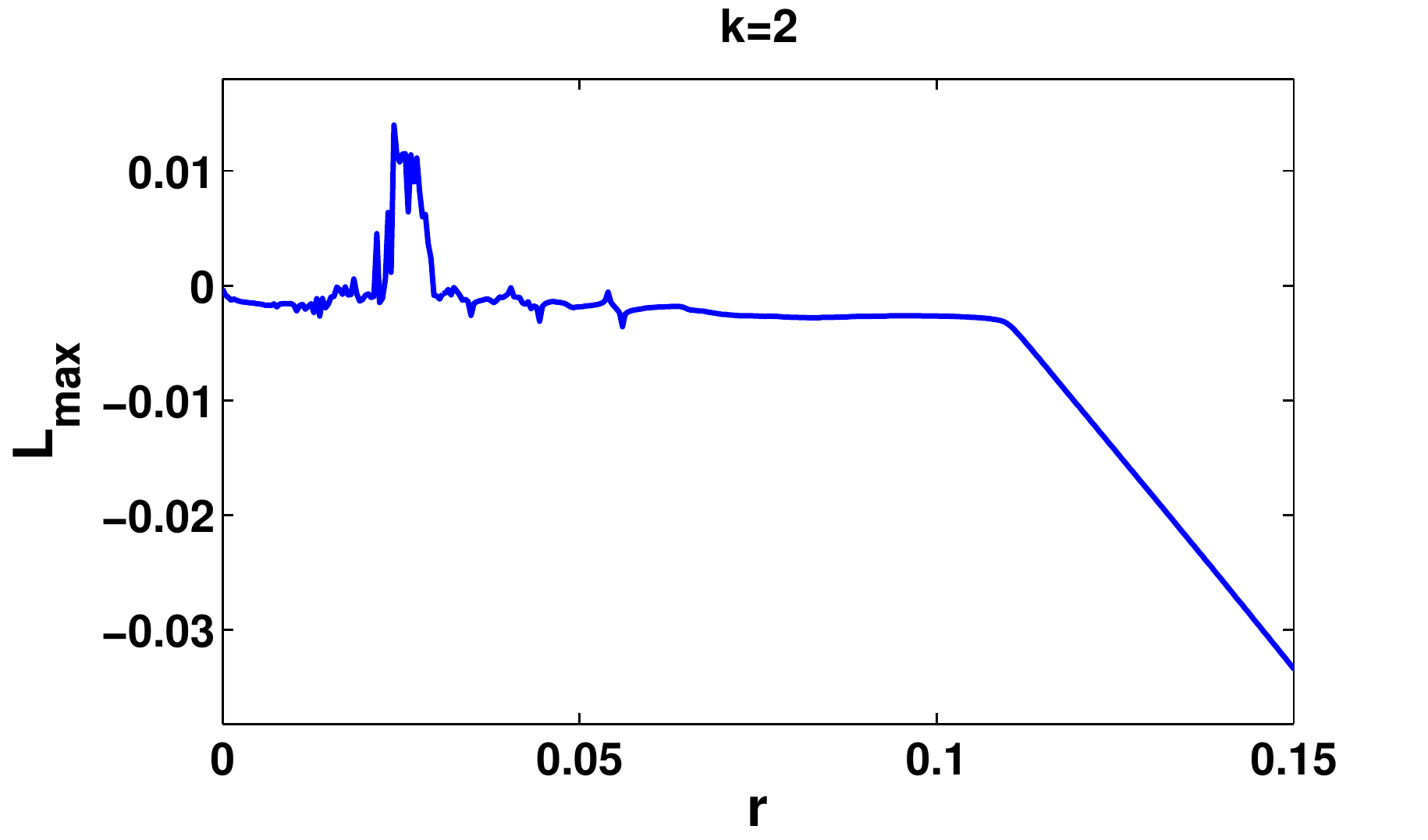}\\[5mm]
\includegraphics[width=7cm,height=4cm]{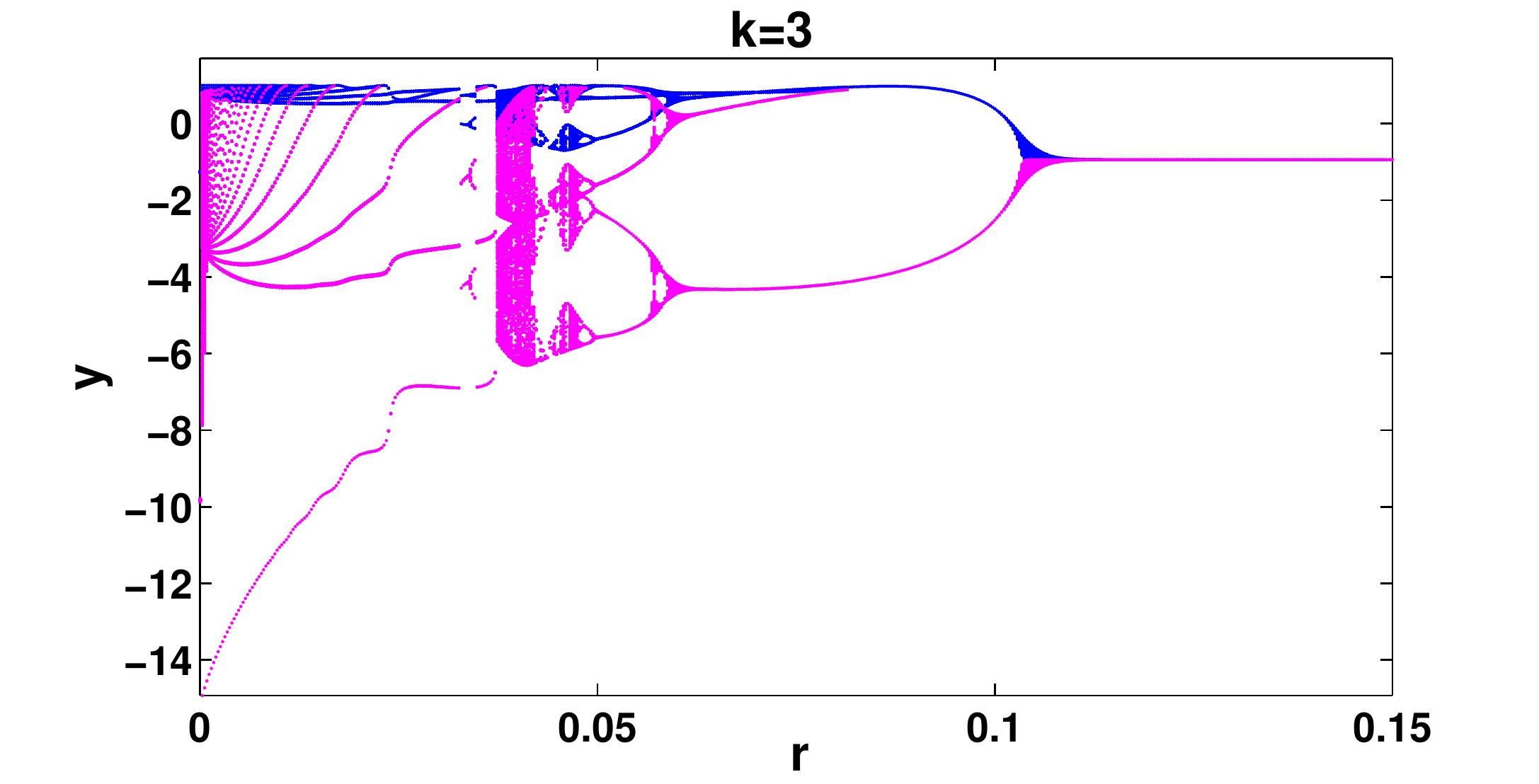}
\includegraphics[width=7cm,height=4cm]{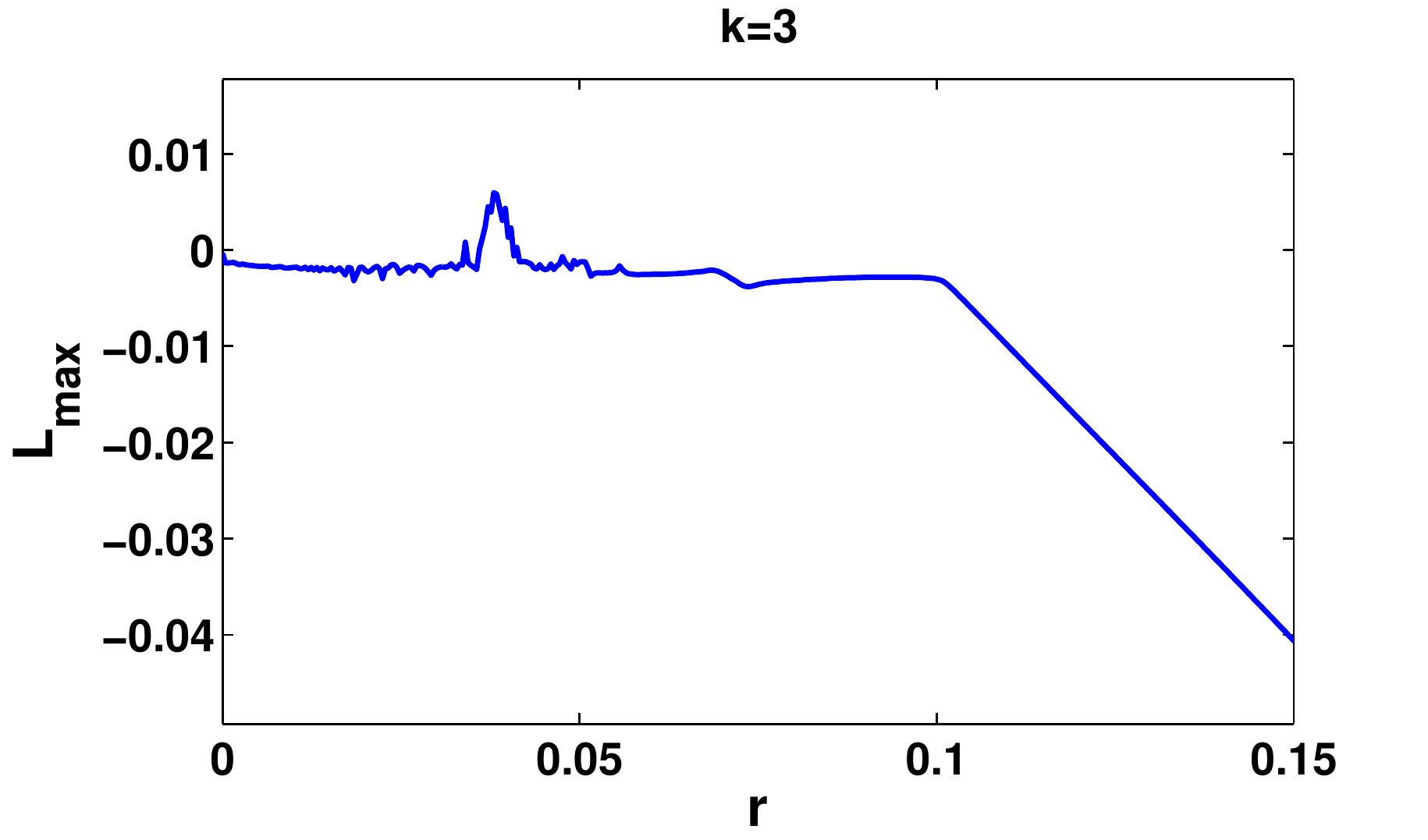}\\[5mm]
\includegraphics[width=7cm,height=4cm]{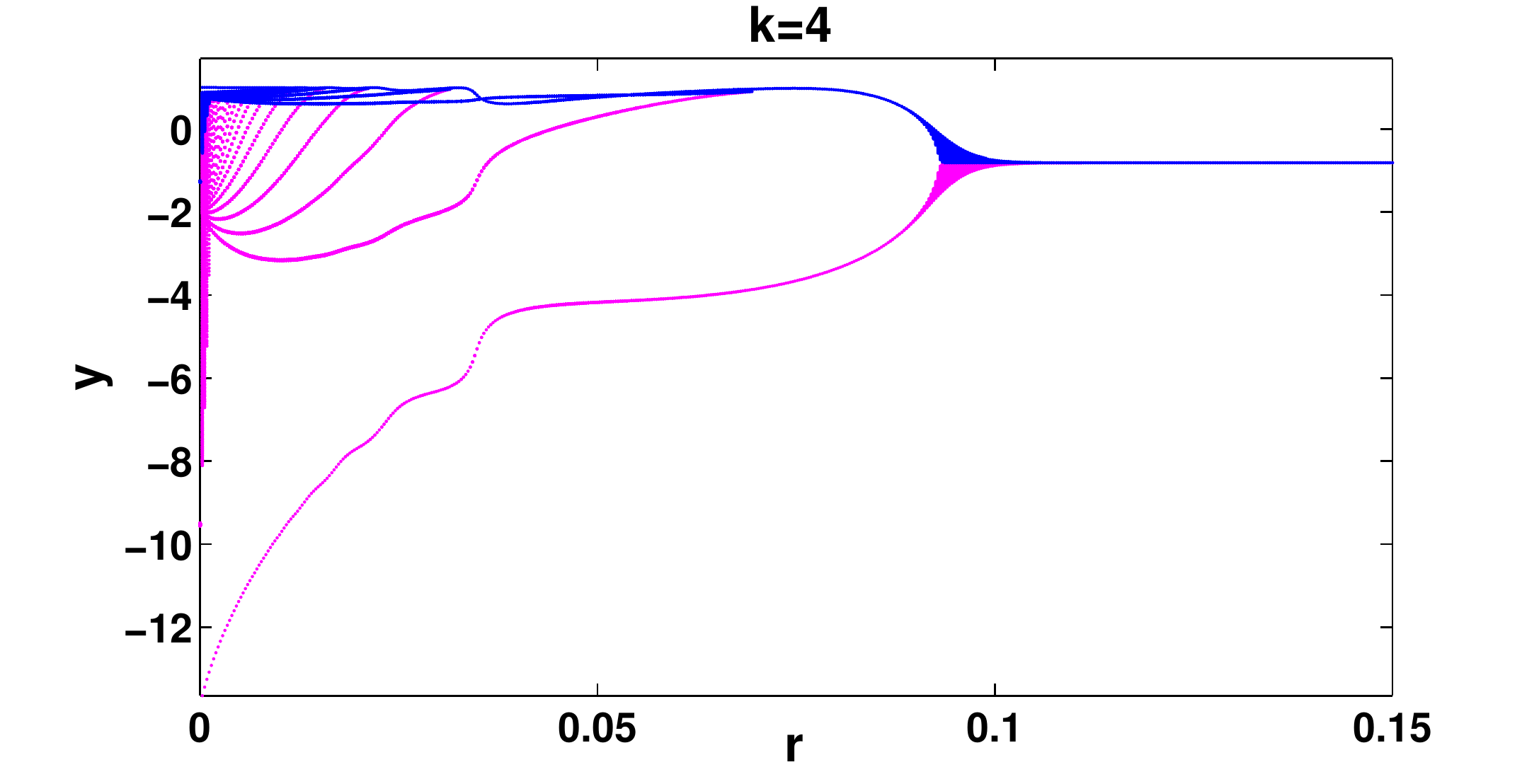}
\includegraphics[width=7cm,height=4cm]{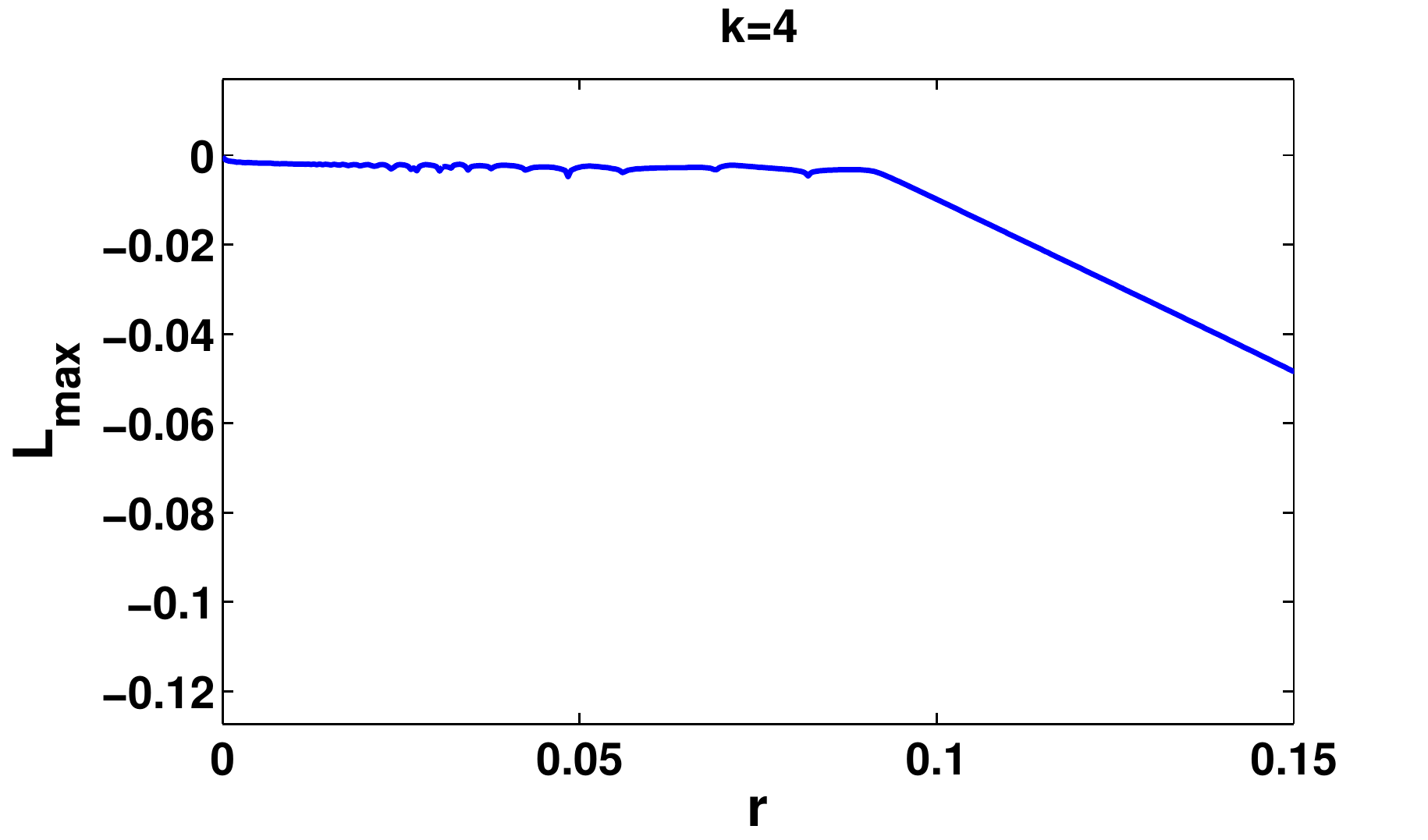}
	\caption{Bifurcation diagram and $L_{max}$ as a function of $r$, for  each value of the intensity of the magnetic flux $k$ indicated in the legends. 
	We used $s=4$, $I_{ext}= 3.25$, as in Fig.~\ref{fig:bifurcation-r}.  
}
	\label{fig:bifurcation-k}
\end{figure}

\end{document}